# The Internet of Things for Smart Manufacturing: A Review

Hui Yang[1*], Soundar Kumara[1], Satish T.S. Bukkapatnam[2], and Fugee Tsung[3]

[1] Harold and Inge Marcus Department of Industrial and Manufacturing Engineering
The Pennsylvania State University, University Park, PA, USA

[2]Department of Industrial and Systems Engineering
Texas A&M University, College Station, TX, USA

[3]Department of Industrial Engineering and Logistics Management
Hong Kong University of Science and Technology, Clear Water Bay, Hong Kong

## Abstract

The modern manufacturing industry is investing in new technologies such as the Internet of Things (IoT), big data analytics, cloud computing and cybersecurity to cope with system complexity, increase information visibility, improve production performance, and gain competitive advantages in the global market. These advances are rapidly enabling a new generation of smart manufacturing, i.e., a cyber-physical system tightly integrating manufacturing enterprises in the physical world with virtual enterprises in cyberspace. To a great extent, realizing the full potential of cyber-physical systems depends on the development of new methodologies on the Internet of Manufacturing Things (IoMT) for data-enabled engineering innovations. This paper presents a review of the IoT technologies and systems that are the drivers and foundations of data-driven innovations in smart manufacturing. We discuss the evolution of internet from computer networks to human networks to the latest era of smart and connected networks of manufacturing things (e.g., materials, sensors, equipment, people, products, and supply chain). In addition, we present a new framework that leverages IoMT and cloud computing to develop a virtual machine network. We further extend our review to IoMT cybersecurity issues that are of paramount importance to businesses and operations, as well as IoT and smart manufacturing policies that are laid out by governments around the world for the future of smart factory. Finally, we present the challenges and opportunities arising from IoMT. We hope this work will help catalyze more in-depth investigations and multi-disciplinary research efforts to advance IoMT technologies.

*Corresponding author:

Hui Yang, e-mail: huy25@psu.edu; voice: (814) 865-7397; Fax: (814) 863-4745

# I. Introduction

The manufacturing sector has a large footprint in the US economy, producing a gross output of $2.2 trillion in 2016, 11.7 % of the total American GDP [1]. To achieve competitive advantages in global markets, modern manufacturing enterprises strive to create new products (or services) with exceptional features such as adaptation, customization, responsiveness, quality and reliability at unprecedented scales. New products have become an integral and indispensable part of everyday life. For example, phones and automobiles are not just communication and travel devices – they are becoming embedded with services which make them act as personal devices. Products are becoming increasingly self-aware. As a result, manufacturing systems are becoming increasingly complex and therefore deploy advanced sensing technologies to increase information visibility and system controllability. Notably, Industry 4.0 is driving manufacturing enterprises to become a new generation of cyber-physical systems towards network-enabled smart manufacturing. The "smartness" level depends to a great extent on data-driven innovations that "*enable all information about the manufacturing process to be available whenever it is needed, wherever it is needed, and in an easily comprehensible form across the enterprise and among interconnected enterprises*" [2, 3]. As smart manufacturing becomes a trend impacting business and economic growth, a large number of networked machines are used increasingly to carry out manufacturing operations. These machines may carry out the same or different functions or tasks, and some machines rely heavily on the output from other machines, e.g., a pipelined product line. The connection between networked machines may also be configured dynamically to increase flexibility and adaptation to customized tasks. As a result, the smart synergy of networked machines is critical to improving the performance of manufacturing systems.

One critical enabling technology for smart manufacturing is the Internet of Things (IoT), which is the formation of a global information network composed of large numbers of interconnected "Things." Here, manufacturing "Things" may include materials, sensors, actuators, controllers, robots, human operators, machines, equipment, products, and material handling equipment to name but a few. The internet-based IoT infrastructure provides an unprecedented opportunity to link manufacturing "Things," services, and applications to achieve effective digital integration of the entire manufacturing enterprise. This integration can be extended from enterprise resource planning (ERP) to supply chain management (SCM) to manufacturing execution system (MES) to process control systems (PCS). However, the rapid growth of large-scale IoT sensing leads to the creation/manifestation of big data that are stored locally or in data repositories distributed over the cloud. *Realizing the full potential of big data for smart manufacturing requires fundamentally new methodologies for large-scale IoT data management, information processing, and manufacturing process control*. For example, the IoT may deploy a multitude of sensors to continuously monitor machine conditions and then transmit data to the cloud. IoT data include not only historical sensor signals and measurements collected from a large number of machines but also on-line data from *in-situ* monitoring of machines. The data can be retrieved easily from the cloud platform to distributed computers for parallel processing and used to extract useful information and prototype algorithms for deployment in the cloud or in the IoT "Things." However, very little has been done to leverage sensing data, known as machine signatures, from a large-scale IoT network of machines to develop new methods and tools for manufacturing systems diagnostics, prognostics, and optimization.

Smart manufacturing goes beyond the automation of manufacturing shop floors but rather depends on data-driven innovations to realize high levels of autonomy and optimization of manufacturing enterprises. As IoT and big data lead to the realization of cyber-physical manufacturing systems, the physical world is reflected in cyberspace through data-driven information processing, modeling and simulation. Analytics in the cyberspace exploit the knowledge and useful information acquired from data to feed optimal actions (or control schemes) back to the physical world. Cyber-physical integration and interaction are indispensable to realizing smart manufacturing. This paper presents a review of IoT technologies and systems that are enablers of data-driven innovations in smart manufacturing. The internet has evolved from hard-wired computer networks through wireless human connected networks to the new era of smart and connected networks of manufacturing things. This trend is integrated with rapid advances in cloud computing, virtual reality, and big data analytics to provide a new paradigm for smart manufacturing. We present a new framework that leverages IoT and cloud computing to develop a virtual machine network. We have also reviewed the IoT cybersecurity issues that are of paramount importance to businesses and operations, as well as IoT and smart manufacturing policies for the future of smart factory defined by governments across the world. Finally, challenges and opportunities in IoMT are described. It is our expectation that this work will catalyze increased multidisciplinary research effort and in-depth investigation to advance the Internet of Manufacturing Things (IoMT) technologies.

The rest of the paper is organized as follows: Section II provides an overview of the Internet of Things. IoT technologies for manufacturing services and applications are discussed and summarized in Section III. Then, we present a case study that leverages IoT and cloud computing to build virtual machine networks in Section IV. IoT cybersecurity issues and manufacturing policies are discussed in Sections V and VI, respectively. The challenges and opportunities to design and develop IoT technologies for smart manufacturing are discussed in Section VII. Finally, we present the conclusions in Section VIII.

# II. IoT Overview

## A. The Evolution of the Internet

The Internet's reach and connectivity have touched every aspect of human endeavor. It is estimated that around 47% of the world population were internet users in 2015 [4]. Fig. 1 shows the evolution from before the internet to the Internet of Things. In the pre-internet stage, telecommunication advanced from the concept of the "speaking telegraph" by Innocenzo Manzetti in 1844 through the first New York to Chicago phone call by Alexander Bell in 1892 to the burgeoning mobile and smart phone technologies. In 1960, the US Department of Defense funded the ARPANET project to develop the first prototype of Internet – interconnected computer networks for fault-tolerant communications. From the 1960s to the 1990s, the world saw rapid developments of content materials in the internet such as emails, information, entertainment, web browsing, and HTML webpages. After the 1990s, the internet began to provide more services to individual users and business users such as online auctions, retailing, shopping, advertisements, search, and financial transactions. Since the 2000s, social networks have facilitated interconnectivity among billions of people, e.g., Linkedin, Facebook, and Twitter. Also, massive open online courses (MOOC) websites are increasingly establishing an internet of students for teaching and education. Most recently, we have witnessed the shift from the internet of people to the internet of things. More and more "smart" devices are connected to the internet. It is estimated that there will be 212 billion "things" connected to the internet by 2020 [5]. The manufacturing industry is also moving towards the new "smart factory," which is envisioned as a cyber-physical system that "*enables all information about the manufacturing process to be available when it is needed, where it is needed, and in the form that it is needed across entire manufacturing supply chains, complete product lifecycles, multiple industries, and small, medium and large enterprises"* [2, 3].

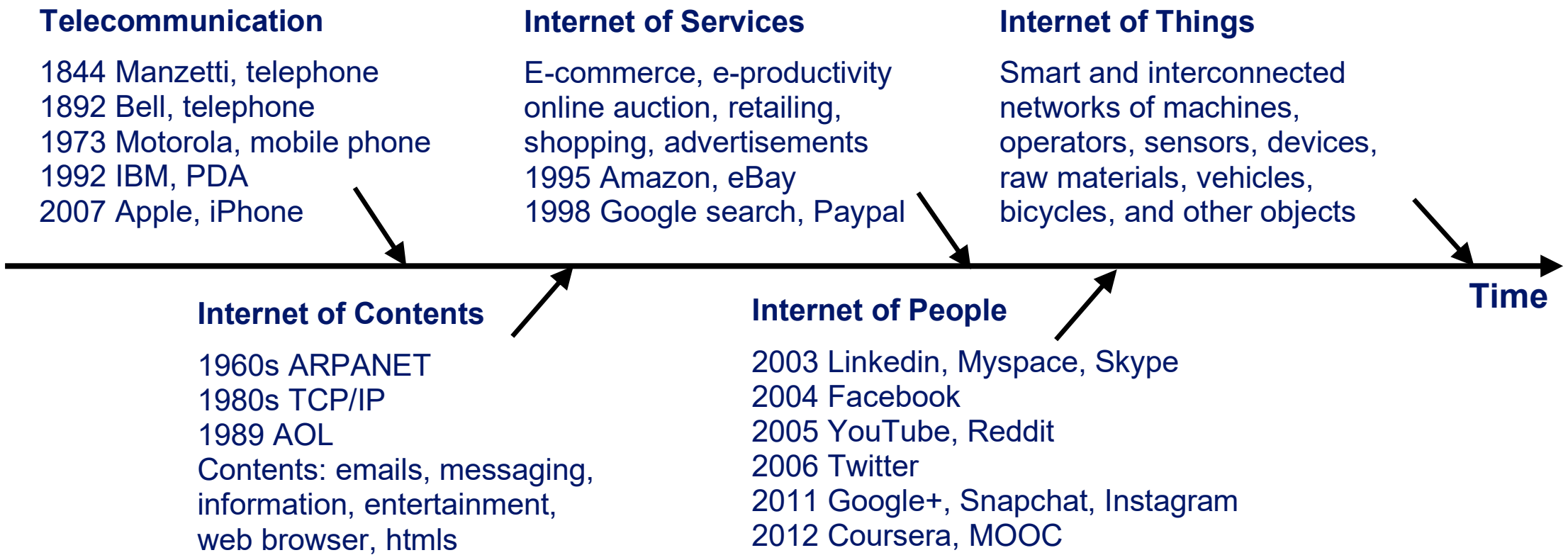

**Fig. 1:** The evolution of the Internet

## B. IoT Sensing

The concept of IoT was first coined by Ashton at the MIT Auto-ID Center in 1999 [6]. The term IoT means the formation of an "Internet" composed of large numbers of interconnected "Things." Here, the "Internet" refers to a global inter-networking infrastructure that uses the TCP/IP protocol to connect and remotely control "Things". High-level communication based on the TCP/IP suite may be supported by a blend of low-level wired and wireless technologies such as Ethernet, Wi-Fi, Bluetooth, ZigBee, radio frequency identification (RFID), or barcodes. "Things" refer to any objects (either physical or virtual) that have unique identities and can sense, collect and/or exchange data about environmental and operational dynamics. Examples of "Things" include vehicles, sensors, actuators, machines, controllers, robots, and human operators. In practice, the IP address and/or a universal unique identifier (UUID) are commonly used to designate a "Thing." This designation greatly enhances the identifiability of "Things," making the integration of "Things" into large-scale IoT networks much easier. The key technologies that integrate "Things" into IoT ecosystems include RFID, wireless sensor networks (WSN), and mobile computing, which are discussed briefly in the following sections:

***RFID:*** RFID technology reads and queries RFID tags attached to an object to automatically identify, monitor, and track the object using radio waves [7]. The basic components of RFID technology are: i) RFID tags, ii) RFID readers, and iii) backend signal processing and IT infrastructure. The RFID tag contains a small microchip that stores data and processes information, as well as an antenna that can receive and transmit data to the reader. RFID tags can be either passive or active. Passive tags harvest energy from the reader's radio waves. Active tags have an embedded power source (e.g., battery) and can operate at a farther distance from the reader. RFID readers transmit an encoded interrogating signal to all tags within range and read out their stored information. Unlike barcodes, the tags do not have to be within the range of sight but only in the range of radio waves. Radio waves provide the energy source for passive tags so that they can respond with their stored identity information. Active RFID sensors often have a longer communication range than passive ones due to the availability of an internal battery. For example, high-frequency active tags (e.g., 3-10GHz) can reach ranges from 300 feet to 1500 feet, while low-frequency passive tags (e.g., 800MHz~900MHz) often operate over ranges between 1 foot and 50 feet. Based on the type of tag and reader, RFID systems are commonly classified into three categories, i.e., Active Reader Passive Tag (ARPT), Active Reader Active Tag (ARAT), and Passive Reader Active Tag (PRAT) [8]. RFID offers a variety of advantages such as low cost, battery-free operation, long range and long lifetime. It is worth mentioning that RFID systems have been used prominently in manufacturing enterprise operations, especially for work-in-process tracking, inventory control, and supply chain visibility management [9].

***Wireless sensor networks (WSN):*** WSNs mainly use spatially distributed autonomous sensors to sense and monitor environmental and operational dynamics of a complex system. Rapid advances in WSNs contribute significantly to the implementation of IoT [10], because "things" are much easier to be connected with each other when many machines are equipped with wireless sensors. Each WSN sensor consists of several components – a radio transceiver to transmit data and receive control signals; a microcontroller providing embedded computing; an analog circuit for signal processing; an embedded operating system; and a power source. Large numbers of WSN sensors are commonly organized into three different types of network topologies, i.e., star, cluster tree, and multi-hop mesh [11]. Because a microcontroller is embedded into sensor nodes to improve the local processing capacity, each individual sensor becomes "smarter" in IoT. Therefore, decision making can be enabled at different levels of an IoT system, i.e., cloud processing, gateway computing, or embedded intelligence in sensor nodes. WSNs have been used widely for civil structure monitoring [10, 12], landslide detection [13], traffic monitoring [14], and machine health monitoring [15, 16]. For example, Bukkapatnam *et al.* installed sensors (i.e., cutting force, vibration, and acoustic emission) to monitor nano-machining dynamics and process-machine interactions to provide higher yields and better repeatability. There are three challenges, i.e., latency, bandwidth and interference, that prevent the ubiquitous application of WSNs in industry. WSNs have a limited bandwidth and update frequency for data transmission. However, it is not necessary to transmit all the raw data through the WSN, but only useful information extracted by the embedded computing. One solution is to transmit features that are extracted from the raw data, and the other is to transmit Fast Fourier Transform (FFT) coefficients (i.e., data compression by Cooley Tukey algorithms) that can be used to reconstruct the raw data.

***Mobile computing:*** Smart phones and tablets bring significant changes in almost every walk of life including the manufacturing industry. Note that smart phones are equipped with internet connectivity, advanced processors, and embedded sensors to obtain acceleration, ambient light, attitude (gyroscope), barometric pressure, GPS location, proximity, and images [17]. As a result, it is easy to integrate mobile computing with IoT systems. For example, IoT things can access the internet or social networks through mobile devices, while IoT sensing capabilities can be enriched by sensors or cameras embedded in the phone. In the past few years, the interplay between IoT systems and mobile phones has significantly increased. The integration of mobile phones with IoT near users promises to improve sensing modalities, increase information-processing capability and also provide better decisions and services in real time.

RFID, WSNs, and mobile computing contribute significantly to the development of IoT sensing systems. IoT sensor nodes are deployed to collect and send data to cloud data centers, while users can control the IoT remotely through the internet. The stored data and analytical results are readily available to users anywhere and at any time using a web-based user interface (e.g., dashboard). As there are different types of IoT sensors, optimal scheduling and planning algorithms for power and computing resources are needed urgently. The existence of heterogeneous sensing networks also requires seamless information exchange and data communication through different protocols to achieve a high level of interoperability.

## C. IoT Data Protocols and Architectures

The efficacy of an IoT system depends to a great extent on the interconnection between many different types of "Things," which may have different communication, processing, storage, and power-supply characteristics. Table I shows a list of 9 data link protocols widely used for data transport in IoT systems. Example protocols used for a short-range and local-area wireless network include Bluetooth, ZigBee, Z-wave, WiFi, and NFC. They are often used to transmit data over short ranges from 10cm to 100 meters. Bluetooth is commonly used for in-vehicle networking and wearable sensing applications [18]. ZigBee is the most popular WSN protocol with low energy consumption well suited for ubiquitous sensing [19]. Z-wave has a very low data rate with a very low energy consumption suitable for smart home and health applications [20]. WiFi is a wireless computer network protocol based on IEEE 802.11 standards, while NFC is commonly seen in contactless payment via smart phones [21]. In addition, there are long-range and wide-area network protocols such as SigFox [22], Neul [23], LoRaWAN [24], and cellular communication technologies. These protocols are commonly used for smart city and environmental applications to transmit data over ranges from 2 kilometers to 200 kilometers.

**Table I**. IoT data link protocols and their characteristics

| Protocol | Standard | Frequency | Range | Data Rates | Applications |
|---|---|---|---|---|---|
| Bluetooth | Bluetooth 4.2 | 2.4GHz | 50-150m | 1Mbps | in-vehicle network<br>wearable sensing<br>smart home |
| ZigBee | IEEE802.15.4 | 2.4GHz | 10-100m | 250kbps | smart home<br>remote control<br>health care |
| Z-Wave | ZAD12837 | 900MHz | 30m | 9.6/40/100kbps | smart home health care |
| WiFi | IEEE 802.11 | 2.4GHz<br>5GHz | 50m | 150~600Mbps | laptops, mobiles, tablets, and digital TVs |
| NFC | ISO/IEC 18000-3 | 13.56MHz | 10cm | 100~420kbps | smartphones, contactless payment |
| Sigfox | Sigfox | 900MHz | 30-50km (Rural)<br>3-10km (Urban) | 10~1000bps | smart city, industrial and environmental applications |
| Neul | Neul | 900MHz | 10km | 10~100kbps | smart city, industrial and environmental applications |
| LoRaWAN | LoRaWAN | Various | 15km (Rural)<br>2-5km (Urban) | 0.3-50 kbps | smart city, industrial and environmental applications |
| Cellular | GSM/GPRS/EDGE (2G), UMTS/HSPA (3G), LTE (4G) | 900 MHz<br>1800 MHz<br>1900 MHz<br>2100MHz | 35km (GSM)<br>200km (HSPA) | 35-170kps(GPRS)<br>120-384kbps(EDGE)<br>384kbps-2Mbps(UMTS)<br>600kbps-10Mbps(HSPA)<br>3-10Mbps (LTE) | cellular networks, mobile phones, and long-distance applications |

The IoT system also uses the internet to connect a large number of "Things." Internet protocol (IP) is a universal standard for data communication over heterogeneous networks. Each "Thing" is assigned a unique IP address. As the number of "Things" connected to the internet is increasing rapidly, scalability of the protocol has emerged as a major challenge. Currently, IPv4 is the 32-bit address system that is on the verge of being incapacitated, i.e., using up all the IP addresses. IPv6 is the new 128-bit address system that has a capacity of approximately $2^{128}$, or $3.4\times10^{38}$ addresses [25]. IPv6 enables every IoT "Thing" to have a unique IP address in the global Internet network. 6LowPAN is a key IPv6-based technology that defines encapsulation and header compression mechanisms independent of the frequency band and physical layers [26]. In other words, 6LowPAN can be used across different communication platforms (e.g., WiFi, ZigBee, 802.15.4), thereby enabling sensors in heterogeneous networks to carry IPv6 packets and become a part of large-scale IoT system.

Specific to manufacturing, MTConnect provides an information model that includes both a common vocabulary (dictionary) and semantics for manufacturing data as well as some communications protocols (specifically through the Agent). MTConnect was developed by the MTConnect Institute to enable manufacturing equipment to communicate data and exchange information using standard Internet technologies, e.g., HTTP and XML (Extensible Mark-Up Language) rather than proprietary formats [27, 28]. MTConnect is a universal protocol for communication between IoT-enabled machines and user-specific applications in the manufacturing shop environment. In other words, open standard grammar and vocabulary are provided by manufacturing dictionary and XML models to define and model manufacturing "Things" such as names, units, values, and contexts of machines and cutting tools. Notably, Table I lists a variety of protocols that can be used to connect and control "Things" remotely. However, MTConnect is a read-only communication protocol that ensures safety by design. In other words, software applications can only request data from MTConnect compatible "Things," but cannot control the machines or equipment through the MTConnect standard.

As shown in Fig. 2, MTConnect consists of three basic components – adapter, agent, and application. The adapter is a software tool that links or converts various data definitions to the MTConnect data definition. Note that the use of an Adapter is the most prevalent means of implementation of the standard, but it is not a requirement. The agent receives data requests from applications and then uses the dictionary and semantics to translate raw data into MTConnect compliant data. Further, MTConnect compliant data will be transmitted to the application for information processing and knowledge discovery, including data requests, storage, analytics, and visualization etc. Examples of applications may include software tools used in manufacturing execution systems (MES), production management systems, enterprise resource planning (ERP), predictive maintenance systems, and visualization dashboards. If the data follow MTConnect definitions, then there will be no need to redefine data for every MTConnect compliant software application. This will help to reduce project costs significantly, optimize production planning, increase manufacturing performance, and improve predictive maintenance.

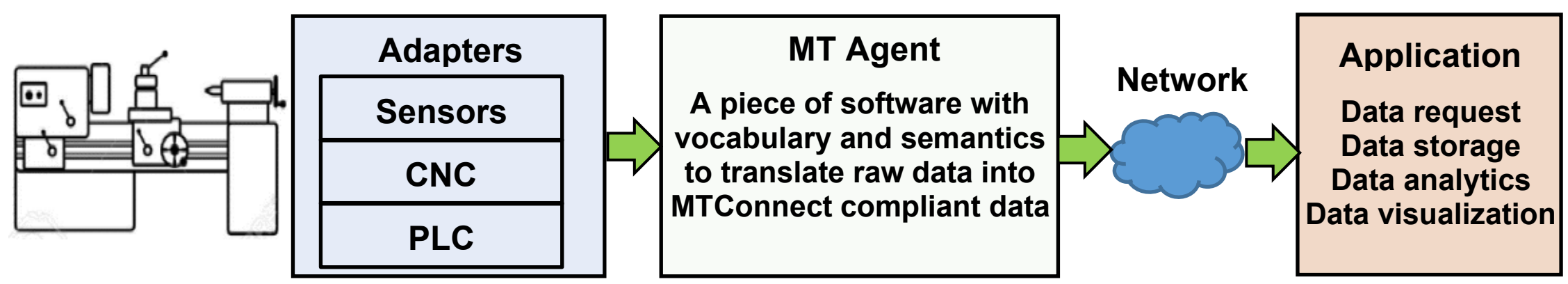

**Fig. 2:** An illustration of the MTConnect standard.

In addition, a number of IoT frameworks and architectures such as RAMI 4.0 and OPC Unified Architecture have been proposed to define the communication structure of Industry 4.0. RAMI 4.0 provides a reference architectural model to define the 3-dimensional map for Industry 4.0. The first dimension is the Factory Hierarchy (i.e., product, field device, control device, station, work center, and enterprise). The second dimension is Architecture (i.e., Asset, Integration, communication, information, function, and business). The third dimension is Product Life Cycle (i.e., from the initial design to the scrapyard). Note that RAMI 4.0 is similar to the Open Systems Interconnection (OSI) model but add two more dimensions that are critical to the industrial systems. Note also that the OSI model uses 7 abstraction layers – physical layer, data link layer, network layer, transport layer, session layer, presentation layer and application layer - to compartmentalize and standardize functions in network communication [29]. As such, the OSI model enables users to communicate over the internet without concern for electrical specifications, binary transmission, or network addressing. Similarly, RAMI 4.0 compartmentalizes and standardizes functions in three different dimensions so as to provide the reference architecture for Industry 4.0. Also, the OPC foundation proposes the OPC Unified Architecture (UA) for data acquisition and information exchange in the RAMI 4.0 framework. Because the same architecture model is used, OPC UA enabled devices and products will speak the same language for effective and efficient communication. However, there are also other IoT architectures currently available such as the IoT standard landscape from NIST, Robot Revolution Initiative (Japan), the Industrial

Internet Consortium (IIC) white paper, Platform Industrie 4.0 white paper, as well as the Cisco white paper. Note that it is difficult for all companies to use the same reference architecture of Industry 4.0 due to competitions in the business world. However, such competition will accelerate the development of a comprehensive IoT framework. As with the first phase of internet development, it is anticipated that competition and collaboration will eventually result in a widely-used IoT framework and architecture for Industry 4.0.

## D. IoT Platforms

Table II shows a list of major IoT platforms and their characteristics. IoT platforms provide the software infrastructure to enable physical "Things" and cyber-world applications to communicate and integrate with each other. Examples of popular platforms include GE Predix, ThingWorx, IBM Watson, Azure, C3 IoT, and AWS. These industrial platforms include a variety of architectural mechanisms including cloud computing, embedded systems, augmented reality integration, data management, software applications, machine learning, and analytical services. Pervasive IoT sensing leads to the proliferation of data. Most IoT platforms provide a service called "dashboard" for data visualization [30]. Currently, dashboard programming has become popular in IoT, because it provides an easy, user-friendly graphical user interface (GUI) to monitor useful key performance indicators (KPIs) quickly and generate reports for decision support. For example, Azure supports a user-configured dashboard that can include a number of resources from the marketplace such as IoT events, time series insights, stream analytics, log analytics, cost analytics, and reports. However, most of these platforms are limited in their ability to fulfill the needs to realize smart manufacturing. In short, these platforms are not specifically designed and customized for the manufacturing industry. It is critical to integrate manufacturing domain expertise with the IoT platforms, which is ultimately required to steer and gain value from the data analysis.

**Table II**. IoT platforms and their characteristics

| Platform | Company | Features |
|---|---|---|
| Predix | GE | ▪ Supports over 60 regulatory frameworks worldwide<br>▪ Pivotal Cloud Foundry<br>▪ Enable industrial-scale Analytics for Asset Performance Management (APM)<br>▪ Cloud platform to build apps for industry |
| ThingWorx | PTC | ▪ Coldlight - IoT Analytics<br>▪ Augmented Reality Integration<br>▪ Machine-to-Machine remote monitoring and service |
| Watson IoT | IBM | ▪ Machine learning and tradeoff Analytics: helps the users to make decisions<br>▪ Visual recognition, Rasberry Pi support<br>▪ Real-Time Insights - Contextualize and analyze real-time IoT data |
| Azure IoT | Microsoft | ▪ Easily integrate Azure IoT Suite with your systems and applications, including Salesforce, SAP, Oracle Database, and Microsoft Dynamics<br>▪ Services: computing, mobile services, data management, and Messaging<br>▪ Enables devices to analyze untapped data automatically |
| AWS IoT | Amazon | ▪ An IoT platform for enterprise application development<br>▪ Supports HTTP, WebSockets, and MQTT<br>▪ Rules Engine can route messages to AWS endpoints<br>▪ Create a virtual model of each device |
| Google IoT Cloud | Google | ▪ Cloud-based platform<br>▪ Modular services: computing, app, query, cloud functions, cloud database<br>▪ Use Google's core infrastructure<br>▪ Committed to open source |
| Machineshop | MachineShop | ▪ Middleware<br>▪ Provides a rich set of different level services<br>▪ Easy integration using industry-standard RESTful API's<br>▪ Edge computing platform |
| Cisco IoT Cloud | Cisco | ▪ Platform as a service (PaaS)<br>▪ REST APIs for send and get data streams<br>▪ Better for tiny IoT prototypes or M2M applications<br>▪ Access to 3$^{rd}$ party APIs |
| Oracle Cloud | Oracle | ▪ Web-based<br>▪ Pre-integrated: Oracle SaaS Auto-Association & Auto-Discovery<br>▪ Rich Connectivity: Cloud & On-premise connectors<br>▪ Recommendations: Built-in recommendation engine for guidance<br>▪ Error Detection & Repair: Alters & Guided Error Handling |

## E. IoT Technologies

There are many enabling technologies (e.g., cloud computing, virtual reality, IPv6, ambient intelligence) contributing to the rapid development and implementation of IoT systems. This section presents the discussion of 3 key technologies – cloud computing, virtual reality, and big data analytics – that promise to improve IoT-enabled manufacturing services.

***Cloud computing:*** Cloud computing provides internet-based computing services, including data storage, data management, KPI computation, data visualization and data analytics amongst others. There are three broad categories of cloud computing services, i.e., infrastructure as a service (IaaS) [31], platform as a service (PaaS) [32], and software as a service (Saas) [33]. IaaS refers to cloud-based services of IT infrastructure such as operating systems, virtual machines, networks, and storage. PaaS provides an environment to develop, test, deploy, and manage IoT software applications. SaaS delivers the services of software applications over the cloud. Cloud computing allows IoT systems to gain ubiquitous access to shared computing and storage resources, thereby overcoming the disadvantage of limited computing resources and storage capability in the "Things." In addition, the integration of cloud computing with IoT offers services such as machine learning and data analytics over the internet, supporting intelligence and decision making in different contexts.

***Virtual Reality (VR) and Augmented Reality (AR):*** The integration of VR and AR with IoT systems is conducive to asset utilization, labor training, root cause diagnosis, and maintenance, among others. VR enables a person's physical presence in the virtual environment and simulates human interactions with virtual objects [34]. VR has been used widely in digital design, workforce training, and predictive maintenance. However, AR augments the real-world, physical environment with computer inputs such as instructions, sound, video, or graphics [35]. AR enables close interaction between the physical world and cyberspace, thereby enhancing the user experience and knowledge about the connection between smart "Things" and the network, human operators, and other "Things." For example, AR is used in the inventory control to check the utilization rate of assets in the storage area of a manufacturing shop. In addition, AR is used by service technicians in the elevator industry to provide remote, predictive and self-guided maintenance and repair services. The use of AR significantly reduces the skill variability between technicians, shortens the repair time, improves the quality of elevator services, and further increases the building efficiency.

***Big data Analytics:*** IoT sensing leads to big data with the following characteristics – high volume, high velocity, high veracity, and high variety [36, 37]. A large number of "Things" generate huge amounts of data in real time. The challenge with manufacturing data is in that it can be "big" in terms of variety and veracity. Variety arises from the diverse data types in manufacturing, from power profiles to machining parameters to acoustic emissions to cutting force signals, each requiring a particular signal acquisition parameter [38]. The manufacturing workshop environment also has a high level of nonstationarity, uncertainty and noise [39]. Veracity is particularly important in the IoT paradigm given the uncertainty (and the lack of quantification of uncertainty) of statistical models. However, the manufacturing industry is not well prepared for changes in the quest for data-driven knowledge [40, 41]. Big data analytics provide efficient and effective methods and tools to handle large-scale IoT data for information processing and manufacturing process control. For example, the new MapReduce framework can be leveraged to develop parallel algorithms for processing massive amounts of data across a distributed cluster of processors or computers and building a virtual machine network [42]. Hadoop is an open software framework for the fast processing of big data and running analytical software on distributed computing clusters [43]. The availability of such big data tools helps to overcome the limited ability of conventional algorithms to process large amounts of data, and further extract useful information and new patterns to help improve the "smartness" level of manufacturing.

# III. Sensor Networks, Manufacturing Services, and Applications

IoT has found applications in many areas such as manufacturing, healthcare, transportation, smart city, and smart home. This section will focus on a review of manufacturing execution systems (MES), sensor-based modeling of manufacturing systems, and the recent development and application of IoT technologies in the manufacturing domain.

## A. Manufacturing Execution Systems

Fig. 3 shows a typical structure of a manufacturing execution system (MES) used in current practice. The objective of MES is to establish transparent data sharing and information exchange between machines, controllers, and the managerial departments in manufacturing shops [44]. At the process levels, there are various proprietary control systems from different vendors such as WSNs, PLCs, and CNC controllers. Gateway computers transmit real-time data streams from control systems in the bottom layer to two database servers. Then, management-level users utilize software applications for process monitoring and data analytics. The MES provides a backbone system for digital performance management, energy management, cost analysis, quality control, and supply chain optimization. Recently, IoT technologies have brought significant changes to the structure of existing MES systems. With the MTConnect protocol and IoT-enabled control systems, MES is moving to cloud platforms. Cloud-based MES systems overcome the difficulty of decoding real-time data streams with proprietary definitions, thereby making data communication, storage, analytics, and reporting much easier to implement.

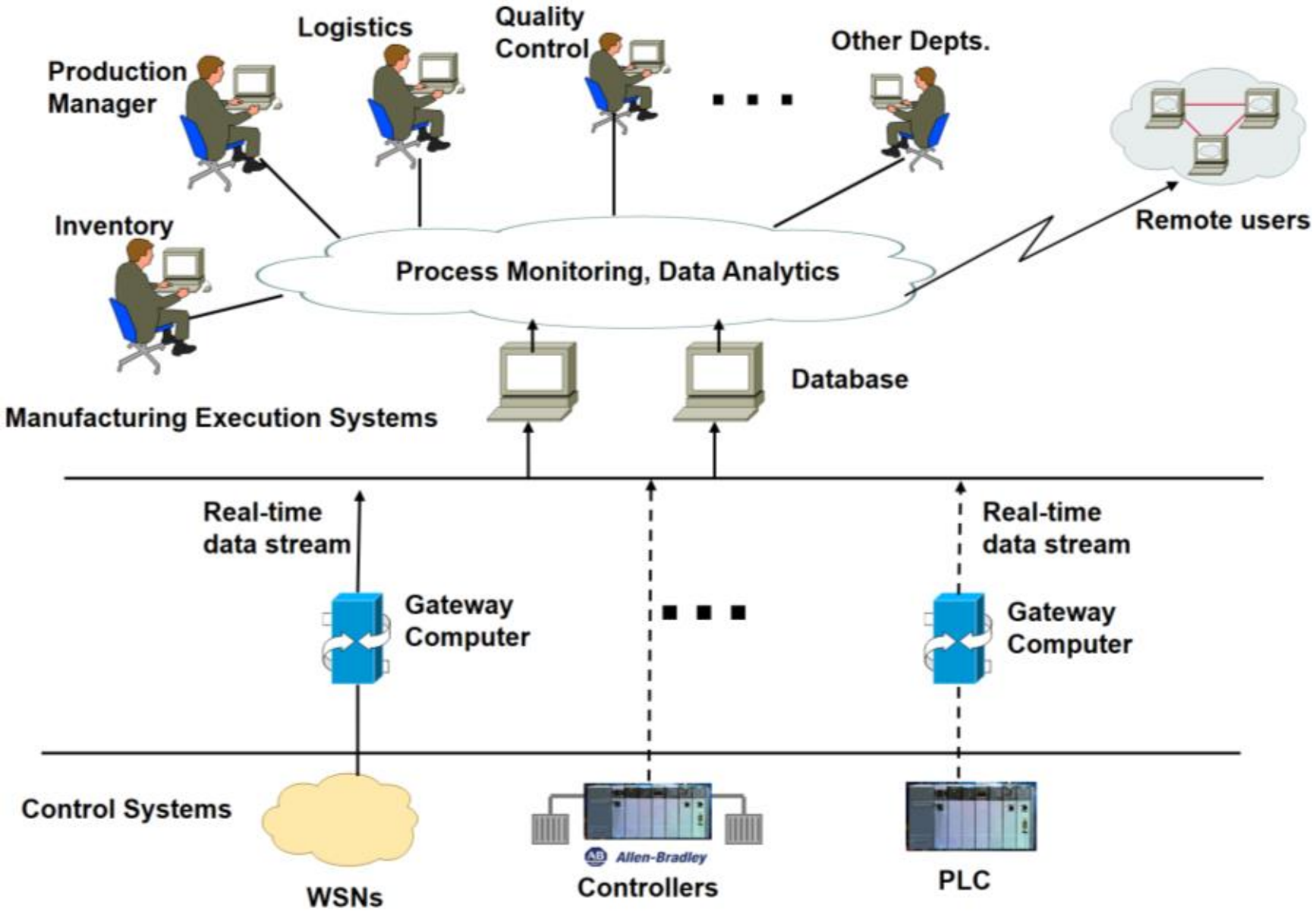

**Fig. 3:** The structure of a manufacturing execution system.

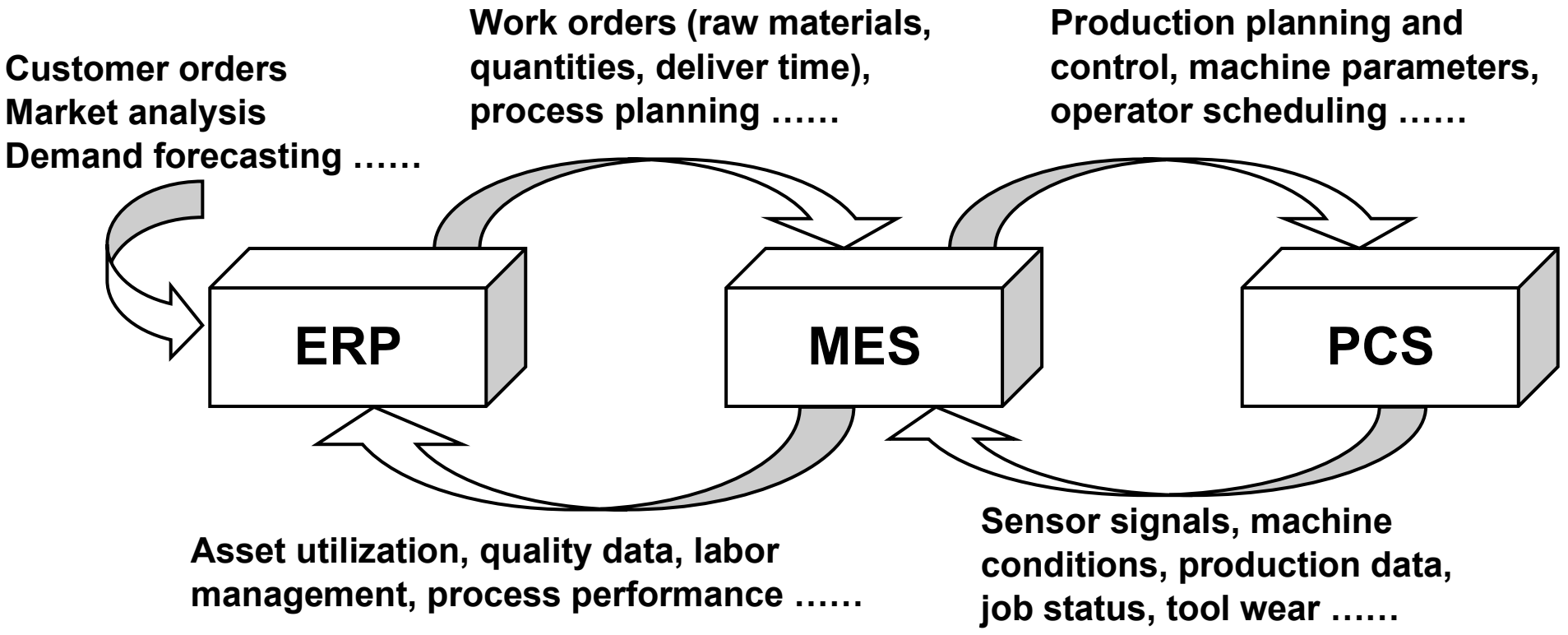

**Fig. 4:** The data flow between ERP, MES, and PCS.

Fig. 4 shows the bidirectional data flow between Enterprise Resource Planning (ERP) systems, MES, and Process Control Systems (PCS), i.e., top down from ERP to PCS and bottom up from PCS to ERP. Fig. 4 follows the Activity models from ISA 95 but focuses more on the data flow. The ERP systems receive inputs of customer orders, market analysis, and demand forecasting [45]. Purchasing and logistics departments will place purchase orders of materials, and also plan, track and monitor shipments. Work orders are then generated and passed to the MES to describe raw materials, order quantities and expected completion time. The MES creates a more detailed plan to complete the production, including the allocation of resources, operator scheduling, and machine parameter settings. When the PCS system is working to fulfill the work orders, in-process data (e.g., real-time sensor signals, machine conditions, production data and job status) will be collected and fed back to the MES. Based on the results of data analytics, the MES adjusts the manufacturing process (e.g., predictive maintenance, operator shifts) to deliver work orders on time. In addition, the MES provides valuable feedback (e.g., asset utilization, quality data, labor management and process performance) to the ERP so that the purchasing department can make changes to the bill of materials. The availability of real-time feedback makes cost analysis, work-in-process predictions, and inventory control more accurate and reliable.

## B. Sensor-based Manufacturing Informatics and Control

Advanced sensing leads to big data populated in ERP, MES, and PCS. Currently, a significant amount of data already exist in the manufacturing domain, but are not fully utilized for real-time process monitoring, fault diagnosis, and performance optimization. Realizing the full potential of MES and advanced sensing depends on the development of new methodologies to extract useful features and patterns from the data, and then exploit the new knowledge to enable smart manufacturing [46]. Here, we categorize sensor-based manufacturing informatics and control into four specific areas as follows:

***Data representation and visualization:*** Sensing systems communicate data in real time with databases (either in the cloud or locally). In many cases, energy budget and bandwidth pose significant challenges on the efficiency and effectiveness of data transmission. For example, battery-supported wireless sensors and active RFID tags commonly face the difficulties in energy budget and bandwidth. As such, a compact representation of data is necessary. For example, Fourier analysis expresses the signals as the summation of sinusoids in different frequency bands. Wavelet representation transforms sensor signals into a combination of orthonormal basis vectors that are locally supported. A compact representation gets rid of the need to store large amounts of raw data, but instead stores significant Fourier or wavelet coefficients for compression and/or transmission purposes. This compact representation also makes the underlying patterns more prominent in the transformed domains so that the extraction of salient features becomes much easier in the context of smart manufacturing [47, 48]. Also, data visualization is critical to presenting key information and patterns to end users in an easily comprehensible way. For example, a customized "Dashboard" GUI can help a user pinpoint critical information of interests, e.g., KPIs, energy usage, machine parameters. Network visualization is also conducive to characterizing and representing the interconnected network of manufacturing "Things," thereby facilitating the formation of a virtual machine network in cyberspace.

***Pattern recognition and feature extraction:*** Data representation and visualization help transform the raw data to alternative domains, e.g., frequency domain, wavelet domain, and state-space domain. The next step is to learn and recognize hidden patterns using pattern recognition methods such as principal component analysis (PCA), data clustering, factor analysis, multilinear subspace learning, and Bayesian networks. Further, feature extraction focuses on the quantification of salient patterns as features for system informatics and control. For examples, Bukkaptnam et al. proposed the wavelet analysis of acoustic emission signals for feature representation in metal cutting [49, 50]. Koh and Shi et al. integrated engineering knowledge with the Haar transformation for tonnage signal analysis and fault detection in stamping processes [51]. Jin and Shi developed feature-preserving data compression of stamping tonnage signals using wavelets [52], and further decomposed press tonnage signals to obtain individual station signals in transfer or progressive die processes [53]. Ding et al. proposed the integration of data-reduction with data-separation tasks for process monitoring and statistical control of waveform signals [54]. Bukkaptnam et al. [47] also proposed an adaptive wavelet method to represent nonlinear dynamic signals for feature extraction in the state space. Bukkaptnam et al. [55, 56] developed local Markov models to predict system dynamics and future evolution in the state space. Yang et al. [57, 58] also developed a new heterogeneous recurrence approach to monitor and control nonlinear stochastic processes. Heterogeneous recurrence analysis was successfully implemented for both sleep apnea monitoring [59] and the identification of dynamic transitions in ultraprecision machining processes [60].

***Sensor data fusion:*** It is common that multiple sensors with different sensitivity to certain operational characteristics are installed in a manufacturing system to collect homogeneous or heterogeneous signals. It may be noted that these multi-sensor signals can be inter-related if they are monitoring system dynamics from different perspectives. Multi-sensor data fusion consists of three critical steps: i) identifying multiscale information flows among multiple sensors; ii) modeling the dynamic evolution of the underlying process dynamics, iii) exploiting the new knowledge from sensor fusion for system informatics and control. Conventionally, linear correlation structures between multiple sensors are characterized to monitor and control manufacturing processes. Effective multi-sensor fusion strategies should consider both information-transfer flows in real-time sensor signals and the evolution of nonlinear dynamics in the underlying processes. For examples, Gang et al. [61] proposed nonlinear coupling analysis of variables by exploiting cross recurrences between them. The nonlinear measure is commonly used in neuroscience to study the interrelationship between neurons. Yang et al. [62, 63] developed a novel wavelet framework - multiscale recurrence analysis - to characterize and quantify the variations of nonlinear dynamics in the underlying processes. Also, Yang et al. [64], and Bukkapatnam and Cheng [65] worked with General Motors to develop local recurrence models to predict the nonlinear and nonstationary evolution of manufacturing operational conditions. Jing and Shi [66] proposed to identify causal relationships from observational data for manufacturing process control. Engineering knowledge was integrated with heuristic rules to learn arc directions in the causal network. Qiang et al. [67] considered nonlinear phase synchronization and thereby physical interactions between correlated functional process variables for conditional monitoring and diagnosis of chemical-mechanical planarization processes.

***Process Control and Decision making***: Once a manufacturing process is out of control, the next step is to take optimal actions to bring the system back under control. The action plan depends on a number of steps such as root cause diagnostics, condition prognostics, and system optimization. Traditional methods for root cause diagnostics include engineering-driven statistical models (e.g., stream of variation analysis, probabilistic graph models) [68, 69] or failure modes and effects analysis (FMEA) [70]. Also, physics-driven models can be formulated based on specific failure mechanisms in the manufacturing system. However, they are often not able to match with real data very well and are therefore inadequate to predict system malfunctions and identify root causes. Data-driven models leverage the real-time sensor signals to characterize and model degradation behaviors in the underlying process. A salient advantage is the ability to transform high-dimensional sensor signals into low-dimensional degradation features for condition prognostics [71, 72].

Further, simulation modeling replicates a real-world manufacturing system, and better explains the underlying mechanisms of the system. Hence, simulation modeling is widely used for diagnostic, prognostic, and optimization purposes. However, discrete-event simulation (DES) tends to track individual entities and their activities in the network of queues. As a result, DES models are not only time-consuming to execute but also provide unrealistic approximations in the setting of mass production or continuous manufacturing. Yang et al. [73] developed continuous-flow simulation models of manufacturing systems using nonlinear differential equations. This approach was used to simulate operational dynamics of a multistage assembly line. The movement of entities is treated as a fluid flow, buffer stocks as water tanks, the conveyor belt as a water pipe and manufacturing stations as valves which control the rates of flow. The continuous-flow models were shown to enable faster and more accurate prediction of aggregate manufacturing performance than DES counterparts. In addition, simulation optimization [74] can be integrated with the wealth of sensor data for manufacturing process modeling and decision support.

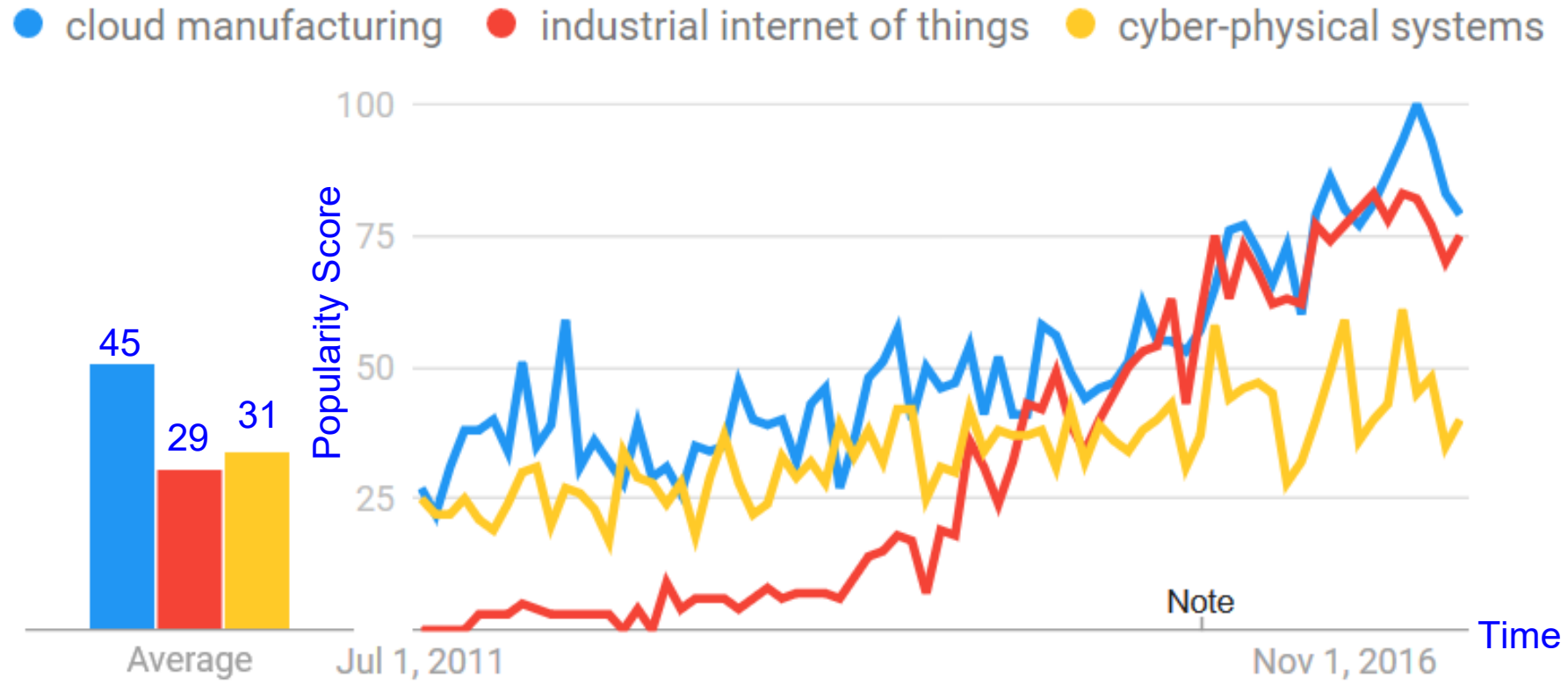

**Fig. 5:** Google trend comparisons of popularity levels of "cloud manufacturing", "industrial internet of things", and "cyber-physical systems" from 07/01/2011 to 08/01/2017. The popularity score represents search interest relative to the highest point on the chart for the given time in the world.

## C. IoT Manufacturing Applications

Fig. 5 shows Google trend comparisons of the popularity levels of "cloud manufacturing", "industrial internet of things", and "cyber-physical systems" from 07/01/2011 to 08/01/2017. The three terms receive increasing attention over the past six years. In particular, industrial IoT yields the fastest increase over the past three years. In this section, we will present a review of IoT manufacturing applications in the following categories: IoT-based cloud manufacturing, cyber-physical manufacturing, energy efficiency management, operations management, safety and ergonomics, as well as supply chain and logistics.

***IoT-based cloud manufacturing***: IoT fuels increasing interests to design and develop new system infrastructures that integrates WSNs and cloud computing into manufacturing settings. For example, Tao et al. [75] developed an architecture of IoT-enabled cloud manufacturing system (i.e., CCIoT-CMfg). This four-layer system provides an opportunity for cloud-based manufacturing service generation, management and applications. Georgakopoulos [76] sketched a road map to harness the power of IoT and cloud computing to enhance manufacturing operations and realize the smart factory. IoT and cloud computing are used to facilitate the real-time monitoring of key plant performance indices, improve productivity, optimally manage inventory level, and improve plant-to-customer traceability. Lin et al. [77] developed a five-stage approach to improve the predictive maintenance of equipment and identify the root causes of yield loss, which is named as advanced manufacturing cloud of things (AMCoT). Zhang et al. [78] proposed an IoT framework for real-time data acquisition and integration, which aims to increase information visibility in the enterprise layer, workshop floor layer, and machine layer for better decisions in manufacturing execution. Internet-based data flow and cloud database in the IoT context effectively facilitate mutual interactions between humans and machines. Cloud computing and analytics can help resolve complex decision-making problems in manufacturing.

***Cyber-physical manufacturing systems:*** The term "cyber-physical manufacturing" is also used in the literature to show the interrelated technologies of IoT, manufacturing and cyber-physical systems. Monostori et al. [79] thoroughly reviewed virtual (i.e., computer science and communication technology) and physical (material science and technology) systems in the field of manufacturing. The authors suggested that cyber-physical manufacturing systems allow adaptive scheduling in production planning, anticipative maintenance strategy, and adaptive production control. Thramboulidis and Christoulakis [80] proposed a UML-based framework (i.e., UML4IoT) to integrate cyber-physical components into the IoT-based manufacturing environment. Such a framework automates the process of generating the IoT-compliant layer allowing both new and legacy cyber-physical components to exploit the IoT connectivity. Tao et al. proposed the IIHub system to support online generation of manufacturing services using encapsulation templates [81]. Particle Swarm Optimization algorithms have also been developed to solve the problem of multi-objective MGrid resource service composition and optimal-selection [82, 83]. Babiceanu and Seker [84] investigated trends in cyber-physical manufacturing systems. They reviewed current applications of virtualization, cloud-based services, and big data analytics in manufacturing settings, and suggested that predictive manufacturing will be an important outcome of the manufacturing cyber-physical system. In addition, Adamson et al. [85] presented the concept of feature-based manufacturing for adaptive equipment control and resource-task matching in a distributed and collaborative manufacturing cyber-physical system.

***Energy efficiency management***: IoT is also utilized for the optimal management of energy efficiency in manufacturing. Qin et al. [86] implemented IoT to optimize energy consumption in additive manufacturing. An IoT-based framework was developed to monitor and analyze energy consumption in the selective laser sintering process and a control system was created to optimize each build and reduce the energy of the entire process. Tan et al. [87] used IoT for the real-time monitoring of energy efficiency on manufacturing shop floors. Energy data were collected and transmitted wirelessly for analysis and feedback, allowing the detection of abnormal energy consumption patterns. The proposed system enables the application of best energy management practice to day-to-day operations. Shaikh et al. [88] investigated enabling technologies to achieve green IoT. Technologies such as RFID, sensor network, and internet were reviewed and their relationship with energy consumption and the environment highlighted. IoT applications were also classified by their impact on the environment. In addition, Tao et al. [89] integrated IoT into the evaluation of energy-saving and emission reduction (ESER). An IoT-enabled system for ESER life cycle assessment was proposed, harnessing the powerful perception ability of IoT for real-time data collection and management. The system facilitates the collection of energy consumption and environmental impact data generated over the entire life cycle of manufacturing, and realizes effective data integration between the ESER evaluation system and the existing enterprise information systems.

***Manufacturing operations management:*** Rymaszewska [90] studied the effect of IoT on the product-service systems of manufacturing industry. Because IoT provides opportunities to access end-users' operations, it helps manufacturing companies to achieve closer and better proximity to customers and change their products accordingly. As such, the IoT-aided system is able to provide the best possible level of service to end users. Li et al. [91] designed an IoT-based predictive maintenance system for equipment used in coal mines. The system incorporates sensors monitoring variables such as vibration and air pressure to collect operational data and transmit them wirelessly to remote servers. Operators can use mobile devices to access the data collected and respond to malfunctions of the equipment. Xu and Chen [92] developed an IoT-based dynamic production scheduling framework for just-in-time manufacturing. The system performs real-time resource status monitoring and dynamic scheduling, helping manufacturers to manipulate production schedules dynamically to maximize production outputs with limited resources. Ding et al. [93] developed an approach to allocate sensors optimally in a multi-station assembly process. By adopting a state-space model and backward-propagation strategy, the distributed sensor system can improve product quality and reduce process downtime. Ding et al. [94] conducted a thorough review of state-of-the-art practices, and investigated the optimal design of distributed sensing systems for quality and productivity improvement.

***Safety and ergonomics***: There are also many research efforts focusing on the design of IoT systems for safety and ergonomics in the manufacturing industry. Boos et al. [95] investigated the use of IoT to address accountability challenges in pharmaceutical manufacturers. Multiple dimensions of accountability (i.e. visibility, responsibility and liability) and control (i.e. transparency, predictability and influence) were studied and a framework was proposed to integrate accountability and control capability in the context of IoT. Sun et al. [96] implemented an IoT-based dam monitoring and pre-alarm system to deal with tailings disposal and prevent the failure of tailing dams. Podgorski et al. [97] designed a conceptual framework for risk management of occupational safety. A framework is proposed for dynamic and personalized occupational risk management, which can continuously assess risks in real time, and monitor the risk level of each worker individually. Environmental and workers' physiological parameters, as well as interactions between workers, the environment and smart physical objects can also be monitored. Guo et al. [98] presented an opportunistic IoT system based on *ad hoc*, opportunistic networking devices using short-range radio techniques such as Wi-Fi and Bluetooth. The system demonstrates an inherent relationship between humans and the opportunistic connections of smart things. It enables information forwarding and dissemination within the opportunistic communities that are formed based on the movement and opportunistic contact of humans. Shirehjini and Semsar [99] developed a mobile 3D user interface to access the IoT-based smart environment. The 3D user interface creates a logical link between physical devices and their virtual representation, allowing users to control the amount and manner in which the IoT automates the environment. In addition, Cheng et al. [100] used nonintrusive real-time worker location sensing and physiological status monitoring technology to monitor the activity (i.e., unsafe behaviors) of construction workers. The proposed system allows the remote monitoring of construction workers' safety performance by fusing their location and physical strain data.

***Supply chain and logistics:*** "Physical Internet" is an IoT-related concept proposed in the domain of manufacturing supply chain and logistics. Meller et al. [101] contributed to the Physical Internet (PI) by developing a road-based PI transit center to efficiently and sustainably transfer trailers from one truck to another. The design of PI transit center was evaluated using key performance indicators. Cheng et al. [102] used complex networks and IoT to address challenges in matching the supply and demand of manufacturing resources. IoT technology was used to realize the intelligent perception and access of various manufacturing resources and capabilities. Reaidy et al [103] proposed an IoT-based platform to fulfill orders in a collaborative warehouse environment. RFID technology was incorporated into an IoT infrastructure to manage decentralized warehouses, improving the competitiveness of warehouses in a dynamic environment and accelerating the adoption of these concepts and technologies in warehouses. Qu et al. [104] developed a dynamic production logistic synchronization to deal with the dynamics of production logistics processes. IoT technology was used to capture the execution dynamics and cloud computing was also incorporated to deal with various dynamics systematically. Fan et al. [105] studied the use of RFID technology to manage inventory inaccuracy in a supply chain. The authors assumed a uniformly distributed demand, and considered factors including fixed investment cost, tag price and shrinkage recovery rate to analyze both RFID and non-RFID cases in both centralized and decentralized supply chains. Qu et al. [104] designed a cost-effective IoT solution for production logistic execution processes with system dynamics. Using sensitivity analysis, optimal IoT solutions were evaluated and analyzed to provide guidance for IoT implementation. Internal and external production logistic processes were combined into an integrated structure to offer a generic system dynamics approach. Hwang et al. [106] employed IoT technology to deal with large fluctuations in demand. An IoT-based performance model was proposed, defining both manufacturing processes and performance indicator formulas. Key Performance Indicators of the overall effectiveness of the equipment were selected to construct an IoT-based production performance model. In addition, Zhou et al. [107] discussed supply chain management in the era of IoT, and provided a review of pertinent papers about business models, architecture for IoT-enabled intelligent decision support systems, the role of IoT technology, and IoT deployment for decision making in production, transport, and service provider selection, and RFID-based inventory management.

In addition to academic research, industrial organizations have increasingly invested in new IoT technologies for process monitoring, operation optimization, fault detection, and optimal control. Table III shows a representative list of companies that implement IoT solutions in industrial case studies. Note that most of examples are for marketing purposes, and more research is urgently needed for IoT system optimization, data modeling, and cybersecurity and so on.

**Table III**. IoT industrial case studies

| Company | Details |
|---|---|
| **Vale Fertilizantes** | Vale used the *GE Predix* platform to improve maintenance strategies and asset reliability, avoiding 25 days of lost production in one year and resulting in a savings of $1.4 million. Corrective maintenance was reduced to zero between 2014 and 2015, and weak acid flow is now above 13 cubic meters per hour. Link: https://www.ge.com/digital/stories/vale-fertilizantes-saves-million-production-losses-asset-performance-management |
| **BMW** | BMW uses *Amazon AWS* for its *car-as-a-sensor (CARASSO)* that collects sensor data to give drivers dynamically updated map information. By running on AWS, CARASSO can adapt to rapidly changing load requirements. By 2018 CARASSO is expected to process data collected by a fleet of 100,000 vehicles traveling more than eight billion kilometers. Link: https://aws.amazon.com/solutions/case-studies/bmw/ |
| **Sandvik Coromant** | Sandvik develops new *predictive analytics on the Microsoft Azure* platform that connects with in-house shop floor control tools to collect the machine data, tool data, and send them to Azure for real-time analysis using machine learning algorithms, as well as process optimization in real-time and the set-up of predictive maintenance schedules and alarms. Link: https://customers.microsoft.com/en-us/story/sandvik-coromant-process-manufacturing-sweden |
| **Toyota Tsusho** | Based on *Amazon AWS*, the company launched a traffic information broadcasting system *TSquare*, which provides users real-time traffic data in Bangkok and 6 suburban provinces. AWS helps process large amounts of traffic data in a scalable and reliable way. Link: https://aws.amazon.com/solutions/case-studies/toyota-tsusho/ |
| **Samsung** | The company developed *S-NET Cloud* based on *Microsoft Azure* for remote energy management of air conditioners. The system saves energy by keeping cooling and heating efficient, using the system air-conditioner sensor, operational data and indoor environmental information. Further, the S-NET system detects equipment malfunctions and performs remote maintenance and management in an integrated manner, using real-time data analytics. Link: https://enterprise.microsoft.com/en-ca/articles/industries/manufacturing-and-resources/remote-energy-management-solution-based-microsoft-azure-iot/ |
| **Cummins Power Generation** | Cummins developed a *Cummins PowerCommand Cloud on Microsoft Azure*, which is a cloud-based remote monitoring solution for generators and power systems. The system can monitor millions of power systems and generators worldwide, thereby improving services, saving lives, and ultimately creating more innovative products that improve quality of life. Link: https://customers.microsoft.com/en-us/story/keeping-the-power-on-when-you-need-it-most |
| **Echelon** | Echelon developed an *adaptive streetlight control system on the IBM Watson IoT platform*. The system boosts energy and operational savings of high-efficiency lighting systems through adaptive lighting control. This helps city managers to take advantage of smart controls that adjust street lighting based on real-time weather data as well as activity levels or time of day. Link: http://news.echelon.com/press-release/corporate/echelon-enables-outdoor-lighting-enhance-public-safety-through-ibm-watson |
| **Marathon Petroleum** | Marathon collects data for analysis on the *GE Predix platform*, and develops collaborative strategy for optimizing the *asset performance management* and optimization. The IoT technology helps Marathon with the service, support, and flexible program design necessary for meeting its ongoing needs. Link: https://www.ge.com/digital/stories/marathon-petroleum-develops-collaborative-strategy-optimizing-apm |
| **Daimler** | Daimler has built a *Detroit Connect* system on *Microsoft Azure* to collect performance data from vehicles on the road and store them in Azure. Fleet managers can view complete fault-event details through the Detroit Connect portal and quickly know when a fault-event has occurred. This helps increase flexibility and reduce costs, and build long-lasting relationships with its customers. Link: https://customers.microsoft.com/en-us/story/daimlertrucks |
| **INNOVYT** | This company developed IoT solutions on *Microsoft Azure* and *Amazon AWS* platforms for real time fleet tracking, alerts and advanced analytics of driving behavior and insights for improving fleet. Link: http://innovyt.com/azure-big-data-solution/ |
| **LightInTheBox** | The company uses *Amazon AWS* to build a highly available website for its customers and save on operating expenses. IoT technology makes it possible to accommodate any transaction, anywhere, and enables the adjustment of computing resources as needed to reduce costs. Link : https://aws.amazon.com/solutions/case-studies/LightInTheBox/ |
| **TraceLink** | This company developed the *Life Science Cloud* platform to ensure compliance throughout the global life science network and global pharmaceutical supply chain. AWS helps the company to fully support the requirements of hundreds of pharmaceutical companies and their partners. Link: https://www.tracelink.com/insights/the-tracelink-life-sciences-cloud-community |

# IV. Case Study - IoT and Cloud Computing to Build Cyber-physical Manufacturing Networks

IoMT integrates sensors, computing units, physical objects (e.g., machines and tools), and services into a network, thereby forming the backbone of a smart manufacturing system. The IoMT network helps a large number of manufacturing "things" to communicate and exchange data. With massive data readily available, IoMT presents an unprecedented opportunity to improve the "smartness" of a manufacturing enterprise.

However, realizing the full potential of IoMT depends on the development of new data-driven methods and tools for smart manufacturing. As IoMT is relatively new, existing methodologies fall short of addressing the internet-like IoMT structures and big data gathered from every corner of a manufacturing enterprise. It is imperative to develop new IoMT analytical methods and tools for smart manufacturing, e.g.

***(i) Data Management***: IoMT communicates large volumes of data at high velocity, calling for new data management techniques (e.g., data access, data structure, data compression, data synthesis, data traceability, data retrieval). It is worth mentioning that there are significant differences between manufacturing data and data from other domains (e.g., computer science, environmental science, healthcare systems). Manufacturing systems involve machines, controllers, robots, sensors, human operators, and elements of other related business units such as inventory, supply chain and management. Data from the network of all manufacturing things show new structures and properties that require efficient handling and storage. Also, data pertinent to specific operations should be efficiently and effectively traced and retrieved to serve the purposes of manufacturing analytics.

***(ii) Information Processing:*** IoMT data contain rich information on fine-grained details of manufacturing systems. There is an urgent need to process the data to extract useful information pertinent to the manufacturing enterprise – from individual machines through networked processes and complete product lifecycles to supply chains. However, data availability does not imply information readiness but requires the development of new information-processing methodologies in the IoMT context. The first stage is data representation to describe the data in alternative domains (e.g., frequency domain, wavelet domain, and state-space domain) so as to reveal hidden information. An effective representation scheme will make statistical measures of salient patterns in the data much simpler in the transformed domain. The second stage is feature extraction to characterize and quantify specific patterns in the IoMT data. Based on the effect sparsity principle [108], there should be a parsimonious set of features sensitive to the state variables to be estimated instead of extraneous noise. Finally, information visualization is necessary to communicate features and patterns efficiently and clearly to end-users through graphics and animations.

***(iii) Decision Making:*** As shown in Fig. 6, IoMT and big data lead to a new generation of cyber-physical manufacturing systems. The physical world is reflected in cyberspace by data-driven information processing, modeling and simulation. Analytics in cyberspace exploits the acquired knowledge and useful information from data to feed optimal actions (or control schemes) back to the physical world. As mentioned above in section II.C, manufacturing decisions of interest include machine monitoring, fault diagnosis, predictive maintenance, inventory optimization, supply chain management, and safety management to name but a few. The "smartness" level in manufacturing depends to a great extent on cyber-physical integration and interaction.

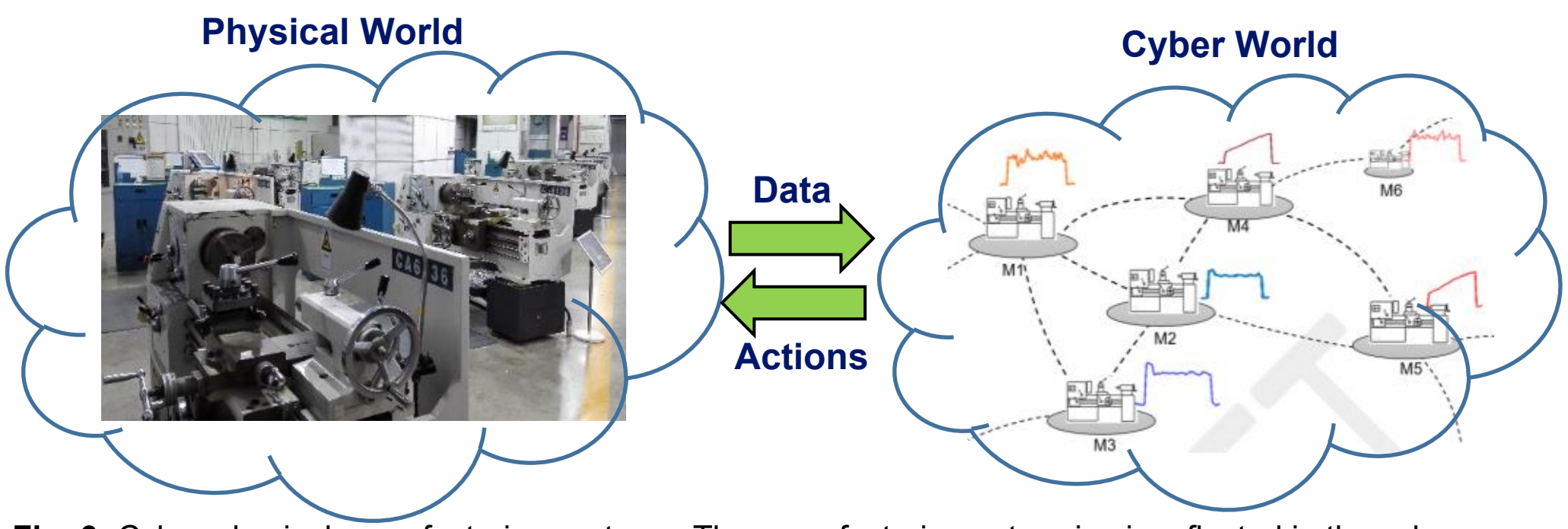

**Fig. 6:** Cyber-physical manufacturing systems. The manufacturing enterprise is reflected in the cyber space through data, and analytics run in the cyber space feed the actions back to the physical world.

In this section, we present a case study of large-scale IoMT machine information processing, network modeling, and condition monitoring. This case study is not comprehensive but serves as an example to leverage the internet-like connection of IoMT machines to build a virtual machine network. As sensor observations contain rich information describing machines' status, this study focuses on the dissimilarity measures between machine signatures (e.g., power profiles from discrete-part manufacturing). Then, each machine is represented as a node of a large-scale network in cyberspace, and node attributes are machine signatures. The edge link and weight depend on the similarity and dissimilarity of node attributes. However, the dimensionality of machine signatures is high and the number of machines is large in the IoMT context. Therefore, we also present an idea of cloud computing for efficient network modeling of large-scale IoMT machines in the cyberspace, which will be detailed in the following subsections.

## IV.A. Physical Machine Networks – Process Monitoring and Control

This case study presents our preliminary studies of stochastic network and parallel algorithms to build a large-scale network of IoMT machines in cyberspace. Notably, most traditional methods focus on the conformance to reference signatures (i.e., "standard" or "normal" ones). However, network models are constructed and optimized using pairwise comparisons of machine profiles. The dissimilarity matrix (consisting of the dissimilarity between each pair of profiles) is obtained from the pairwise comparison, rather than from a column in a reference comparison. For conventional reference comparisons, the computational workload is low, and easy to implement. The difference against the reference profile can be directly used as an indicator to determine if the current profile is normal or not. However, it is necessary to empirically and/or statistically establish a "normal" signature from a historical record of profiles. On the contrary, network modeling does not need to establish a "normal" signature but rather leverages the pairwise dissimilarity information to automatically group large numbers of profiles into homogeneous clusters. As such, the proposed network approach will provide a better representation of information in the data and further offers opportunities for visual analysis of machine conditions.

The proposed machine-network models are generally applicable to monitoring P2P dynamics in manufacturing processes. In other words, one machine repeatedly manufactures the same type of discrete part in large quantities (*the high volume, low mix scenario*). Further, P2P network models can be used for different types of parts. For example, if there are two different kinds of part, then power profiles from the same part will have a higher level of similarity than those from different parts. This will lead to another application of product classification - group parts into homogeneous clusters. Network visualization will provide categorization of parts, evaluation of energy consumption, and further help production planning. For *the low volume high mix scenario*, network models can be potentially applicable to product classification or detection of process characteristics (e.g., types of materials, machining procedures, and specific tools used). For example, M2M networks can help to extract useful information about machine utilization, power usage, and condition monitoring, which will help further optimize factory operations, reducing equipment downtime and maintenance costs. Virtual machine networks have great potential to shift current manufacturing practices towards globalized production optimization and management. IoMT energy management provides a major opportunity to optimize the energy consumption and realize green and sustainable manufacturing.

## IV.B. Virtual Machine Networks

Virtual manufacturing overcomes many practical limitations in the physical world and provides a greater level of flexibility to optimize a variety of manufacturing actions (e.g., production planning, quality control, maintenance scheduling) in cyberspace. As manufacturing is highly complex and involves multifarious elements, there are potentially several types of virtual manufacturing networks including (a) machine networks; (b) supply chain networks; (c) human resource networks; and (d) customer networks to name a few. In this present study, we focus on the development of virtual machine networks. It may be noted that social networks are essentially an internet of people, and people can communicate with each other easily through a network. However, it is easy to build a virtual machine network but difficult to enable communication between networked machines. Here, we propose to build virtual machine-to-machine networks by allowing each machine to exchange real-time attributes with each other (e.g., machine signatures, profiles, events). As such, machines can form a community or a group in the network that collectively provides a subnetwork of machines with similar attributes.

For example, power profiles are a machine signature that describes the energy consumption of successive operations in a discrete-manufacturing process. Fig. 6 (right) shows that IoMT-enabled machines communicate power profiles with a distinct morphology and pattern during the cutting phase. Some of them show nominal patterns (e.g., M2 and M3), whereas others have larger variations (e.g., M1 and M6) and elevated patterns (e.g., M4 and M5). Note that machine signatures may vary due to a number of factors such as the product, machine type, procedure, and anomalies. In the large-scale IoMT context, each machine can communicate its attributes (e.g., power profiles) for every discrete part produced, thereby allowing the quantification of both *part-to-part (P2P)* and *machine-to-machine (M2M) dissimilarities* in the attributes.

Such an IoMT-based virtual machine network provides great opportunities for ***i) Condition monitoring and quality control*:** Machines with similar conditions can be grouped into the same cluster. The structure of a virtual machine network not only provides useful information on the machine status and utilization statistics but also offers the opportunity of profile-based machine clustering, product categorization, and online quality control. ***ii) Planning and scheduling***: The structure of a virtual machine network varies dynamically because machine profiles change over time. Such a dynamic network can further help optimize maintenance decisions, manufacturing planning, and scheduling. For example, we can proactively assign a machine's workload to other (normal) machines and schedule maintenance, when a machine is moving towards the "machine failure" cluster in the network. ***ii) Smart manufacturing***: For a large-scale manufacturing system, advanced sensing increases information visibility and helps cope with high-level complexity in the system. IoMT provides an opportunity to realize the virtual machine network for smart manufacturing. For example, machines communicate with each other to report their status and exchange information for optimal planning and scheduling. This will substantially help to create value from data, optimize factory operations and reduce maintenance costs and equipment downtime.

## IV.C. Network Modeling and Analytics

Advanced sensing in the large-scale IoMT context communicates rich data streams. As shown in Fig. 6, IoMT connects a large number of machines in the manufacturing system and generates overwhelmingly big data. For an individual machine, power profiles can be collected during the production of discrete parts. When a large number of parts are produced, the IoMT will generate tens of thousands of power profiles. P2P variations in power profiles provide a wealth of information pertinent to machine conditions and production performance. This will enable engineers to make proactive decisions to adjust processes and maintain machines, improving the quality of products and reducing the re-work rate. For a group of machines, IoMT sensing provides an unprecedented opportunity to embody machines in a large-scale network to enable smart manufacturing. However, the number of machines and data volume pose significant challenges for the construction and optimization of a cyber-physical machine network. There is an urgent need to extract pertinent knowledge about manufacturing operations (i.e., from one machine to a group of machines) and then exploit the knowledge acquired for decision making. Realizing the promise of IoMT depends to a greater extent on information-processing capability. Little has been done to address the fundamental issues important to big data analytics in the large-scale IoMT context. In this case study, we propose to develop virtual machine network models from the following perspectives:

***(1) Customized P2P network***: It is not uncommon for IoMT sensing to collect long-term monitoring data from an individual machine. As shown in Fig. 7, during the production of a part, the signal waveform changes significantly in different segments (i.e., different stages of the manufacturing operation). Between two different parts, the signals are similar to each other but with variations. Therefore, we propose to develop a network model of stochastic ***P2P*** dynamics for customized monitoring of machine conditions, where each part is represented as a network node and the node attributes are profile data for this part.

***(2) Population M2M network***: There are also similarities and dissimilarities in profile patterns between two different machines. Therefore, we propose to develop a ***virtual M2M*** network model, where each node represents an individual machine and node attributes are the dominant profile patterns or aggregated properties. The choice of node attributes is highly dependent on domain-specific applications. Such an M2M network will help engineers and managers to identify machine communities that share similar operational conditions, study machine variations within each community, and pinpoint an individual machine in one of the communities for monitoring and maintenance purposes.

***(3) Parallel graph analytics***: However, such network modeling is computationally expensive due to the population size and data volume. Traditional serial-computing schemes are limited in their ability to represent networks efficiently and provide real-time analytics in the IoMT setting. Note that the power of IoMT lies in the inclusion of more machines to form a network topology, links, and communities. Hence, we propose to develop parallel algorithms for efficient network modeling and optimization of the large-scale IoMT, as well as further develop network-based predictive analytics for smart manufacturing.

In the following four subsections, we will discuss the technical steps towards the construction and optimization of virtual machine networks (i.e., pattern matching, network modeling, predictive analytics, and parallel computing). These steps are not meant to be comprehensive or exclusive, but rather serve as initial ideas for IoMT network modeling.

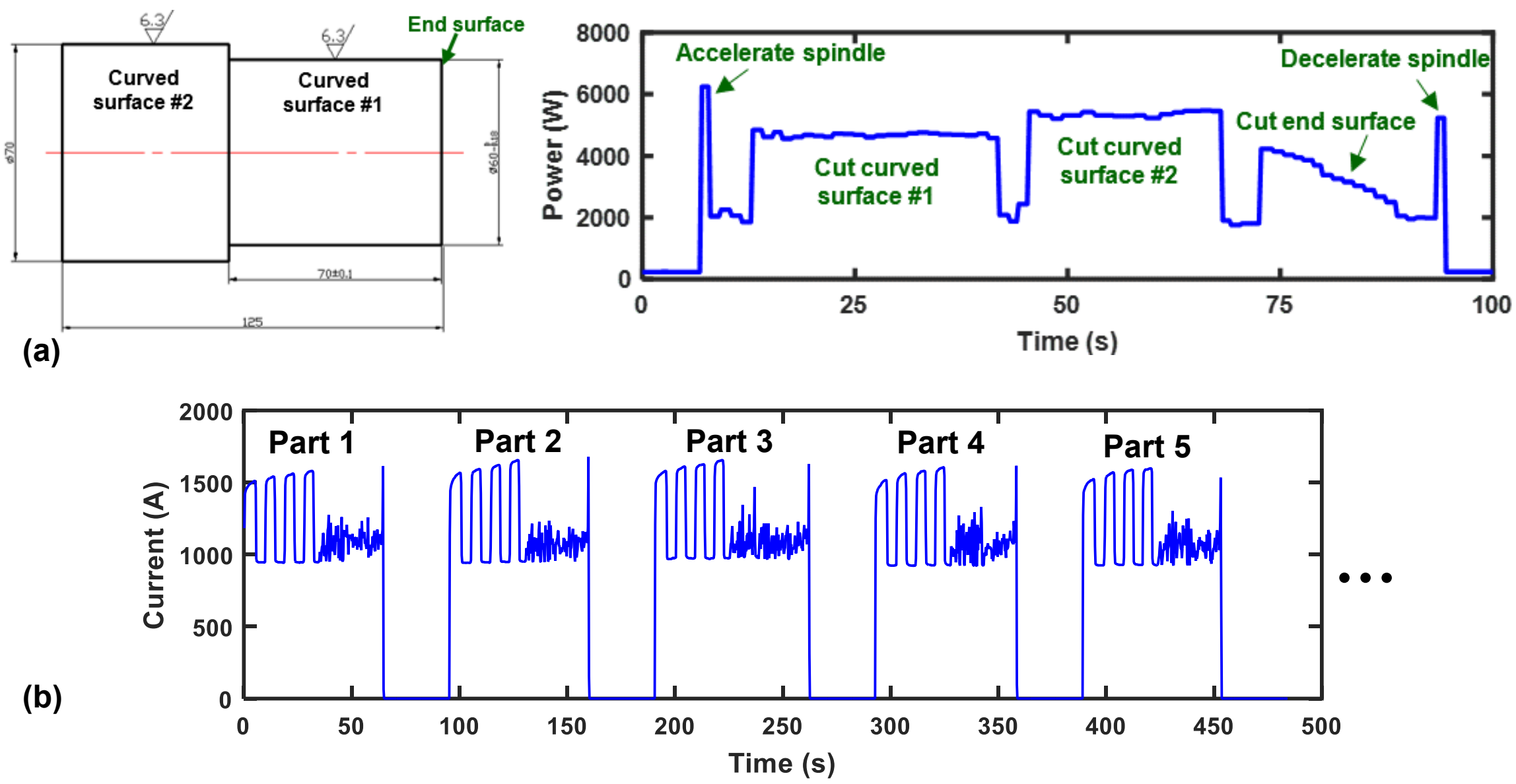

**Fig. 7:** (a) The CAD file and power profiles from the machining operation; (b) P2P variations in current profiles when a welding machine produces parts with the same design.

## *IV.C.1 Pattern Matching*

Fig. 7a shows the CAD file and power profiles for a machining operation. The variation in energy consumption can be due to machine parameters (e.g., rotations per minute, depth of cut, and feed rate), tool conditions, materials properties, and other uncertainty factors. Fig. 7b shows the part-to-part (P2P) power profiles when a welding machine cyclically produces discrete parts with the same design. Although this investigation uses power consumption data as an illustration, there may be other profiles of interest such as acoustic emission, cutting force, or vibration. Note that profile patterns are similar to each other because the parts have the same design but have variations (i.e., due to machine and process variations). Because profile patterns are very pertinent to process dynamics, pattern matching will provide a good opportunity to monitor the condition of machines and tools. Fig. 7 shows there are pattern variations between the power profiles from Part 1 to Part 5, although they are all of the same design. Conventional methods focus on the comparison between the current profile and reference profiles (i.e., "standard" or "normal" ones). Here, we propose to perform a pairwise comparison of machine profiles using either P2P or M2M network methods. Note that a dissimilarity matrix (that is, the dissimilarity between each pair of profiles) is obtained from pairwise comparison, rather than only being a column in reference comparison. However, two profiles can be misaligned due to discrete sampling and phase shift. For e.g., Parts 1-5 in Fig. 7 show a typical pattern, but there are variations in shape, amplitude, and phase. This poses significant challenges to the characterization and quantification of pattern interrelationships (i.e., similarity and dissimilarity) between profiles.

In the literature, such interrelationships are estimated by methods such as correlation and mutual information. Note that correlation is a second-order quantity evaluating merely the linear dependency between two profiles $x_1(t)$ and $x_2(t)$. It should be noted that linear correlation cannot capture the nonlinear interdependence between variables adequately. Mutual information $MI_{x_1,x_2}$ characterizes and quantifies both linear and nonlinear correlations but requires stationarity in the computation, i.e.,

$$MI_{x_1,x_2} = \sum_{x_2}\sum_{x_1} Pr(x_1, x_2)\log\left(\frac{Pr(x_1, x_2)}{Pr(x_1)Pr(x_2)}\right) \quad (1)$$

Yun et al. [109] developed an information theoretic approach that used mutual information to measure the nonlinear correlation between variables (i.e., analogous to profiles) for variable clustering and predictive modeling. Gang et al. [61] proposed nonlinear coupling analysis of variables by exploiting cross recurrences between them. This nonlinear measure is commonly used in neuroscience to study the interrelationship between neurons. In addition, Zhou et al. [110] investigated the discrete wavelet transform of cycle-based profiles and developed a wavelet control chart for process monitoring.

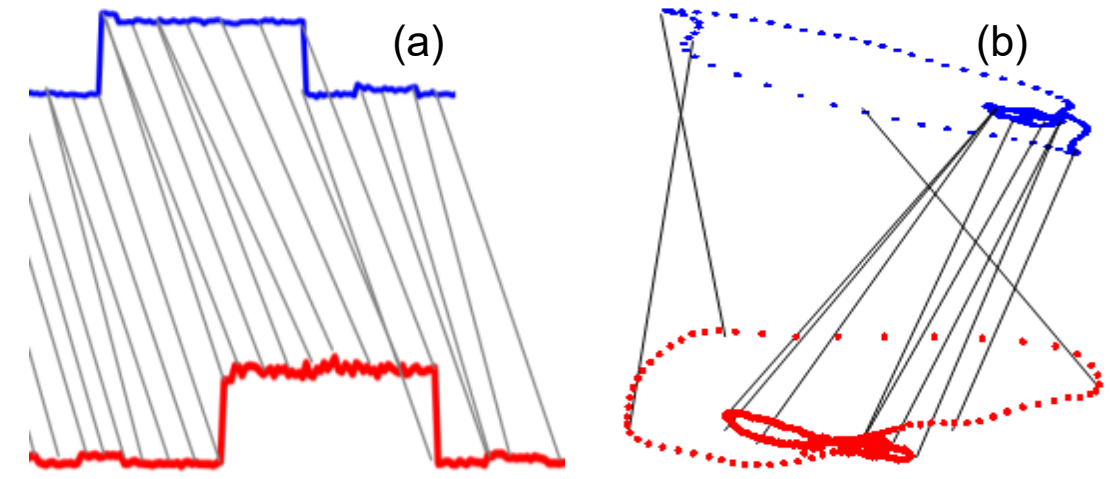

**Fig. 8:** One-dimensional (a) and three-dimensional dynamic time warping for pattern matching

In order to measure the morphologic dissimilarity between profile data, Yang et al. [111] proposed one-dimensional and multi-dimensional dynamic time warping (see Fig. 8). Note that profile alignment is imperative to measure pattern dissimilarity. If we do not use the warping approach and measure the difference between part profiles directly, this will contaminate useful information and will not yield meaningful results in most cases. However, dynamic time warping aligns two signatures optimally and yields meaningful results by comparing the morphology of corresponding segments. Given two profiles $\overrightarrow{\boldsymbol{x}_1}(t)$, $t$ = 1, 2, …, $n_1$ and $\overrightarrow{\boldsymbol{x}_2}(t)$, $t$ = 1, 2, …, $n_2$, the dissimilarity between $\overrightarrow{\boldsymbol{x}_1}(t)$ and $\overrightarrow{\boldsymbol{x}_2}(t)$ is then measured as $\sum_{(t_i,t_j)\in\zeta}\left\|\overrightarrow{\boldsymbol{x}_1}(t_i) - \overrightarrow{\boldsymbol{x}_2}(t_j)\right\|$. To find the optimal warping path $\zeta$, a dynamic programming algorithm iteratively searches:

$$d(i,j) = min\begin{pmatrix} d(i,j-1) + w(i,j) \\ d(i-1,j-1) + w(i,j) \\ d(i-1,j) + w(i,j) \end{pmatrix} \qquad (2)$$

where the initial condition is $d(1,1) = w(1,1) = \|\overrightarrow{\boldsymbol{x}_1}(t_1) - \overrightarrow{\boldsymbol{x}_2}(t_1)\|$ and a window size constraint is $|i-j| < r$ . The normalized dissimilarity between $\overrightarrow{\boldsymbol{x}_1}(t)$ and $\overrightarrow{\boldsymbol{x}_2}(t)$ are obtained as $\Delta(\overrightarrow{\boldsymbol{x}_1},\overrightarrow{\boldsymbol{x}_2}) = d(n_1,n_2)/(n_1+n_2)$. As a result, machine profiles are optimally aligned for the measurement of pattern dissimilarities. If pattern matching is performed for every pair of profiles, then a warping matrix will be generated to provide pairwise similarity and dissimilarity among profiles.

### *IV.C.2 Network Modeling*

Although the warping matrix contains rich information about the variations in machine profiles (i.e., either P2P or M2M), it is difficult to use the matrix itself as a predictor in predictive models for manufacturing applications. There is an urgent need to develop novel methods and tools that will enable and assist the exploitation of dissimilarity matrices to make optimal decisions in manufacturing. Because these machines are networked elements in the manufacturing system, it is natural to use network theory to provide analytical methods to study the interrelationship and interactions between machines. The nodes or vertices of such networks will be machines and the edges or links will be interactions (i.e., similarity or dissimilarity in profiles) between machines.

The next step is to optimally represent each P2P (or M2M) machine profile as a network node in a high-dimensional space. The distance between nodes should preserve the dissimilarity between two corresponding profiles. Fig. 9 illustrates the network modeling of six machine profiles. A dissimilarity matrix provides pertinent information about the variations of machine signatures. By optimizing the location of nodes in the network, node-to-node distances preserve the profile-to-profile dissimilarities in the warping matrix of Fig. 9a. For example, Fig. 9 shows that dissimilarities between M1 and others (M2-M6) are preserved as the Euclidean distance between node M1 and others. Let $\boldsymbol{s}_i$ and $\boldsymbol{s}_j$ denote the location of $i^{th}$ and $j^{th}$ nodes in the network and $\delta_{ij}$ is the dissimilarity between $i^{th}$ and $j^{th}$ machine profiles in the warping matrix $\Delta$. Then, the objective function of network modeling can be formulated as:

$$\min \sum_{i<j} (\|\boldsymbol{s}_i - \boldsymbol{s}_j\| - \delta_{ij})\,;\; i,j \in [1,n] \qquad (2)$$

where $\|\cdot\|$ can be the Euclidean norm or other distance measures, depending on the specific application. This approach represents each power profile as a network node based on the pairwise dissimilarity measures, which greatly reduces the dimensionality of the data and thereby identifies the "best data" to represent the machine's condition. In the presence of a small number of machines (or profiles), optimizing the locations of network nodes $\boldsymbol{s}_i$'s can be achieved by existing algorithms such as multi-dimensional scaling (MDS) [112] and scaling by majorizing a complicated function (SMACOF) [113]. However, the computational complexity of MDS is $O(N^3)$ because of the centering operation and eigen-decomposition. SMACOF utilizes the Guttman transform that consists of a large matrix multiplication. Existing algorithms tend to have limitations in addressing the emerging challenges in network modeling of a large-scale IoMT network.

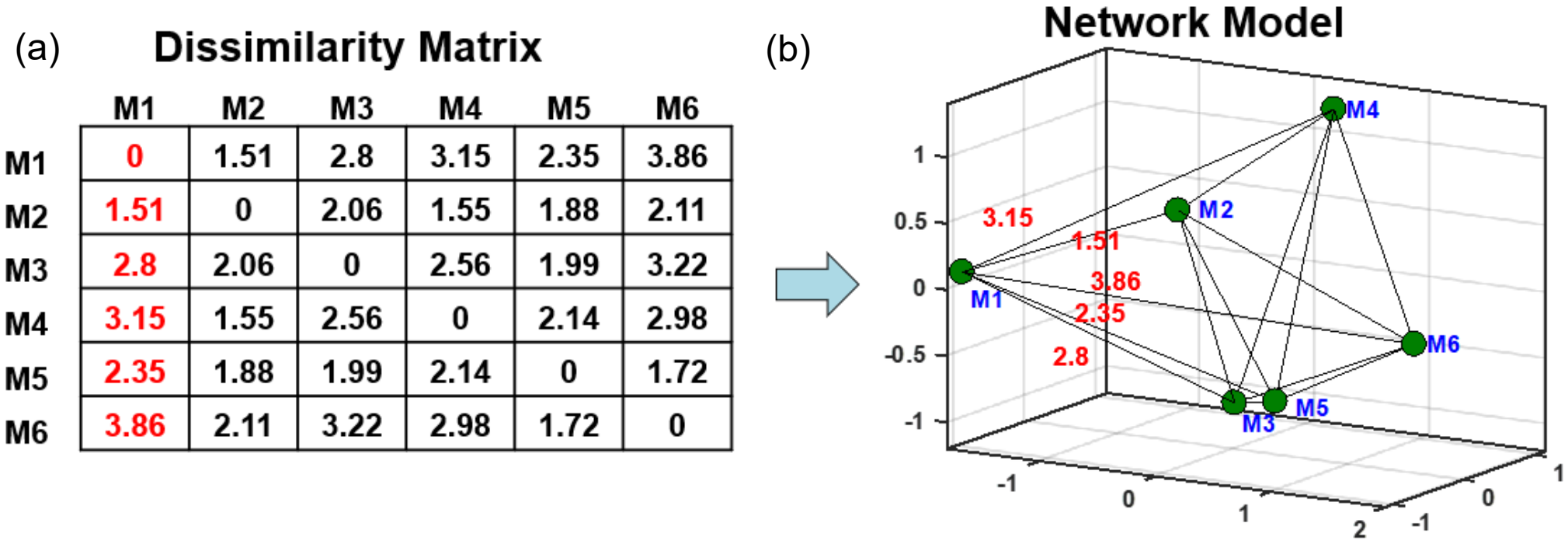

| | M1 | M2 | M3 | M4 | M5 | M6 |
|---|---|---|---|---|---|---|
| M1 | 0 | 1.51 | 2.8 | 3.15 | 2.35 | 3.86 |
| M2 | 1.51 | 0 | 2.06 | 1.55 | 1.88 | 2.11 |
| M3 | 2.8 | 2.06 | 0 | 2.56 | 1.99 | 3.22 |
| M4 | 3.15 | 1.55 | 2.56 | 0 | 2.14 | 2.98 |
| M5 | 2.35 | 1.88 | 1.99 | 2.14 | 0 | 1.72 |
| M6 | 3.86 | 2.11 | 3.22 | 2.98 | 1.72 | 0 |

**Fig. 9:** (a) Dissimilarity matrix of six machine profiles; (b) A network model with node-to-node distances preserving the profile-to-profile dissimilarity matrix in (a).

### *IV.C.3 Cloud Computing*

Because serial algorithms often lead to prohibitive computation time in large-scale IoMT, there is an urgent need to scale up the algorithm and use large-scale machine learning in cloud computing to complete the optimization task collaboratively. Parallel algorithms pipeline the overall computing task into multiple computers (or processors) for collaborative processing. As shown in Fig. 10, each computer carries out part of the computation and works simultaneously with other computers to combine results to build virtual machine networks, reducing the computing time significantly. Nowadays, the availability of multi-core CPUs, cell processors (e.g., GPUs), and cloud computing make parallel algorithms easily implementable with off-the-shelf strategies such as multi-threading and single-instruction-multiple-data.

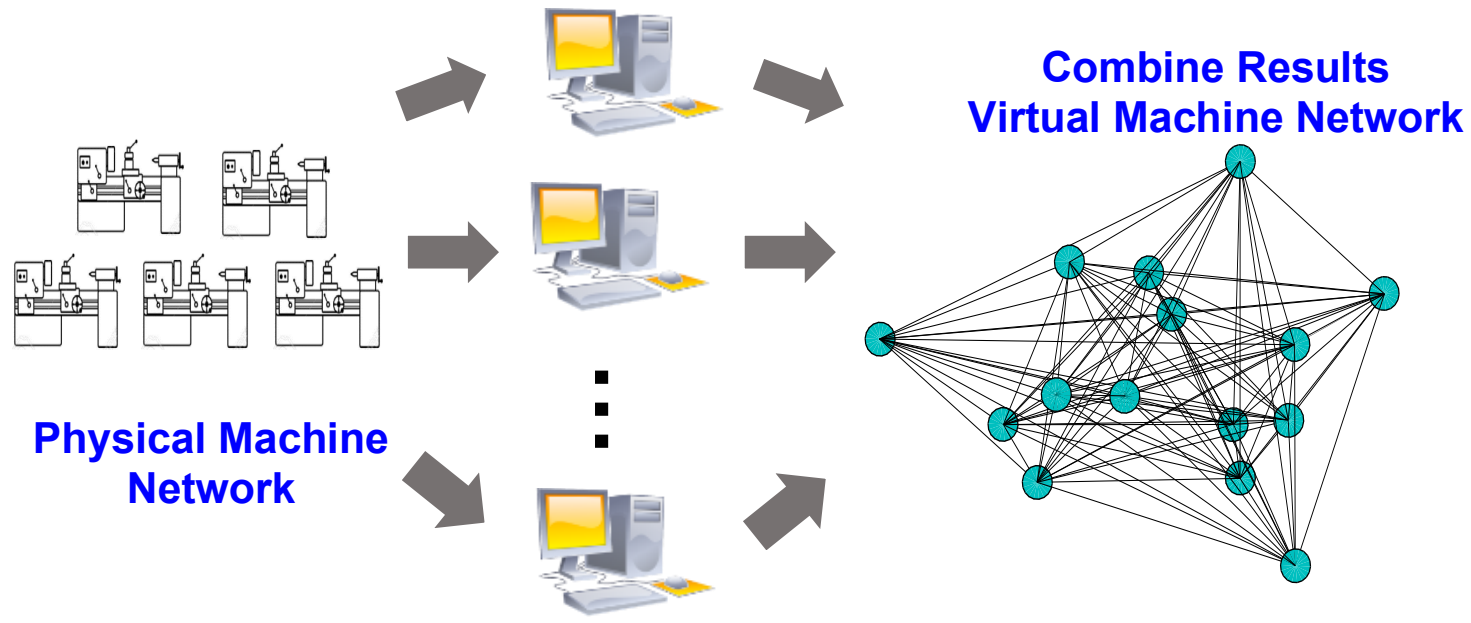

**Fig. 10:** Map reduce and cloud computing to optimize the virtual machine network

Due to the high dimensionality of IoMT data, second-order algorithms are often difficult and even prohibitive. It is preferable to use first-order approaches such as gradient-based algorithms. Also, large volumes of data pose significant challenges to optimization algorithms as each iteration must process the entire data set. The stochastic gradient approach is well suited to handle IoMT data because each iteration or subtask processes a limited subset of samples. For convex functions, stochastic gradient approaches are shown to guarantee convergence in most cases. Therefore, Kan et al. [114] proposed to integrate stochastic gradient algorithms with batch learning, i.e., mini-batch stochastic network algorithms, to model virtual machine networks. As opposed to considering one sample at a time, mini-batch stochastic network algorithms handle multiple samples (i.e., a mini-batch) simultaneously in each iteration. See more details in [114]. Once the virtual machine network is constructed, changes in machine conditions are encoded as network dynamics. We will discuss IoMT network analytics in more detail in the next subsection.

### *IV.C.4 Predictive Analytics*

Virtual machine networks provide a new paradigm for exploring future physical spaces and perform predictive analytics towards an anticipatory manufacturing enterprise. Network analytics are generally applicable to provide decision support for future production and market variations. Here, virtual machine networks provide a new means of studying the manufacturing system using a network structure and topology.

Node attributes (e.g., coordinates of nodes), once optimized, can be used as features for condition monitoring and predictive modeling. For example, if P2P variations of a single machine are monitored in the IoMT system, each node will represent the operational profile of one discrete part. If the node is located away from the cluster of normal condition in the virtual machine network, maintenance services need to be scheduled to prevent machine failure. For a group of machines (i.e., M2M networks), each node represents a machine or its signature (e.g., power profiles, features, or patterns). Predictive models can be constructed to predict the machine's condition based on its node attributes. If node attributes show the machine condition has a high probability of being abnormal, engineers can re-assign its jobs to other machines and schedule maintenance. In the state of the art, there are various modeling approaches available for predictive applications, including linear regression models [115], neural networks [116], self-organizing networks [117, 118] and particle filtering [119] Practitioners can select an optimal predictive modeling approach based on the complexity of data and requirements of processing speed.

Machines are also interconnected, and each machine has its nominal profile patterns and temporal variations. However, abnormal events in one machine may cause a cascade of follow-up events on neighboring machines in the network. In the social network, sociologists show that social connections and interactions have significant impacts on a person's behavior [120, 121]. Network edges (or links) represent the connections between machines, thereby providing important information about the behaviors of manufacturing systems. It is worth investigating whether the organization of links in large-scale IoMT networks are random or follow specific principles for unique properties or orders. The link and topological patterns facilitate the detection of community structures. There will be small variations in node properties within the same community but large variations between two different communities. Virtual machine networks provide a graphical representation of large numbers of machines and also groups machines with similar conditions into homogeneous clusters. This enables condition monitoring of machines using visual analytics of the data. For example, we can pinpoint each profile from a machine in the network clusters for monitoring or classification purposes. Raghavan et al. [122] developed a label propagation algorithm (LPA) that is a near linear time algorithm to effectively and efficiently identify community structures in large-scale IoMT networks. LPA is widely used and is a part of software packages like R, Python, Java, and iGraph libraries. As natural networks are often uncontrolled and exhibit self-organizing behavior, self-organized M2M networks may be investigated further to increase robustness against external and internal uncertainty in manufacturing.

Further, network topology has been shown to influence the performance of processes such as event formation, information diffusion, navigation, search and others. Topological measures that are widely used to exploit meaningful information in networked processes include node degree, link density, average path length, network diameter and clustering coefficient. A comprehensive review of network topological measures can be found in [123, 124]. Although topological measures are important, they may be insufficient to describe specific functionalities of the machine networks. There is a need to unearth patterns of node attribute, link organization, community structure, and network topology from the large-scale IoMT network. For examples, are there link and topological patterns we can exploit to optimize the design of facilities? Which community does the machine belong to? Are there manufacturing jobs that need to be re-assigned? Are there preventive maintenance services that can be delivered based on real-time machine conditions?

# V. IoT and Cybersecurity in Manufacturing

With the rapid advance of IOT, it is anticipated that the manufacturing industry will see more and more IoT-based devices, applications and services in the next few years. As manufacturing equipment is a part of the critical infrastructure for economic growth, they can easily become the target of malicious attackers. The interconnection of IoT devices, cloud databases, and information networks makes the IoT system vulnerable to cyber-attacks. Therefore, IoT cybersecurity is of primary concern in smart manufacturing. It is imperative to develop new cybersecurity frameworks and methodologies that will help facilitate the widespread adoption of IoMT. MTConnect advocates a read-only option when the upper-level MES interacts with the smart manufacturing "Things" in the IoT system [27, 28]. In other words, software applications can only read data from the network of sensors, machines, controllers in the lower-level PCS system, but cannot write data to control or damage the manufacturing infrastructure.

As shown in Fig. 11, the National Institute of Standards and Technology (NIST) developed the cybersecurity framework (CSF) for manufacturing implementation [125]. This CSF includes five critical components, i.e., $i$) Identify: What processes and assets need protection? $ii$) Protect: What safeguards are available? $iii$) Detect: What techniques can identify incidents? $iv$) Respond: What techniques can contain the impact of incidents? $v$) Recover: What techniques can restore capabilities? This framework can be used to measure the performance of different cybersecurity solutions, thereby helping further improve the implementation of IoT and cybersecurity systems in manufacturing environments.

In the past few years, cybersecurity has fueled increasing interest in the manufacturing community. For examples, Hutchins et al. [126] proposed a framework to identify vulnerabilities in automotive manufacturing systems, which considers the data flows within a manufacturing enterprise and throughout the supply chain. Desmit et al. [127] proposed an intersection mapping approach to identify cyber-physical vulnerabilities and predict their influence on intelligent manufacturing systems. Sturm et al. [128] focused on cyber-physical vulnerabilities in additive manufacturing (AM), and made the following recommendations to improve the AM cybersecurity, i.e., improving software checks; hashing/securing signing/blockchain; improving process monitoring; operator training.

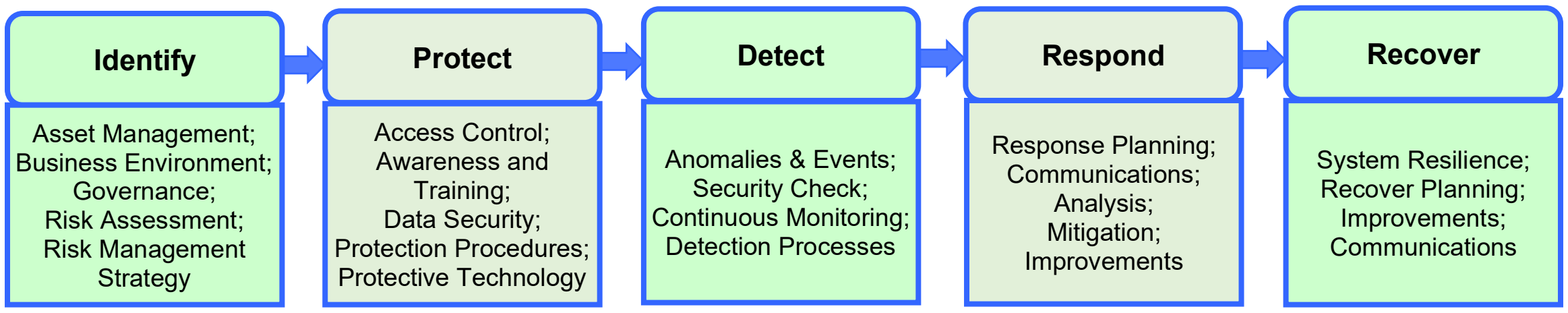

**Fig. 11:** The cybersecurity framework for manufacturing implementation from NIST.

Furthermore, there are a number of innovative techniques to protect the security and privacy of IoMT systems including cryptographic solutions, intrusion identification, and blockchain technology.

***Cryptographic solutions***: Duan et al. [129] designed a data-centric access control framework to provide secure access to smart-grid services in a publish/subscribe model. Seo et al. [130] focused on the development of lightweight key management protocols for scalable and distributed authentication. Zhang et al. [131] propose a password-authenticated group key exchange protocol and prove its security in a standard model, which does not require short passwords to be pre-shared among users. Saxena et al. [132] designed a lightweight authentication and key agreement protocol for the long-term evolution (LTE) network to support secure and efficient communications between IoT devices and their users. Note that manufacturing data can be encrypted locally and in the cloud using the PGP standard [using entropy generated keys and AES encryption] as well as transmitted through communications encrypted on the chip by hardware.

***Intrusion Identification:*** Siboni et al. [133] developed a cybersecurity testbed framework that allows wearable device designers and manufacturers to evaluate the security of the devices in a simulated environment. Saeed et al. [134] used random neural networks to develop an intrusion detection and prevention scheme for IoT systems. Vincent et al. [135] was inspired by side-channel schemes used to detect Trojans in integrated circuits, and then proposed a product/process design approach to enable real-time attack detection, i.e., changes to a manufactured part's intrinsic behavior. Thames et al. [136] developed a cyberattack detection algorithm based on ensemble learning with neural networks, and further integrated response mechanisms into the cloud-based manufacturing architecture.

***Blockchain Technology:*** As a new approach to decentralized computation and assets management in the BitCoin system, the blockchain technology [137, 138] has promised to help to address scalability and security challenges in IoMT. Ghuli et al. [139] proposed a decentralized system to register and assign IoT devices to an owner based on the blockchain technology. Bahga et al. [140] developed a decentralized, peer-to-peer platform to implement IoT systems based on the blockchain technology. This platform will enable users in a decentralized, trustless, peer-to-peer network to interact with each other without the need for a trusted intermediary so as to improve the cybersecurity of IoT systems.

# VI. IoT Manufacturing Policies and Strategies

This section briefly discusses IoT manufacturing policies and strategies from various countries and industrial organizations. Such policies and strategies are the main drivers for the development and practical implementation of IoT technologies, and play important roles in pushing the paradigm shift towards smart manufacturing in the next few decades. Currently, there are many policy initiatives across the world aiming to promote smart manufacturing and stimulate economic growth.

***USA***: The PCAST* report in 2012 identified Advanced Manufacturing as a path towards to revitalizing U.S. leadership in manufacturing, creating high-quality jobs, and ensuring national security [141]. Next-generation manufacturing is envisioned to depend on the effective use and coordination of automation, sensing, networking, data, information, and computation. The goal is to enable high-rate, cost-effective, repeatable production for practical industrial implementation. In particular, advanced sensing, networking and process control are identified as key technology areas for smart manufacturing. In the past few years, smart manufacturing has attracted significant interest. To build a robust, sustainable R&D infrastructure, Manufacturing USA - formerly known as National Network for Manufacturing Innovation - has established several networked Manufacturing Innovation Institutes as follows:

- AFFOA (Advanced Functional Fabrics of America): http://www.rle.mit.edu/fabric/
- AIM Photonics (American Institute for Manufacturing Integrated Photonics): http://www.aimphotonics.com/
- America Makes: https://americamakes.us/
- ARM (Advanced Robotics Manufacturing): http://www.arminstitute.org/
- ARMI (Advanced Regenerative Manufacturing Institute): http://www.armiusa.org/
- CESMII (Clean Energy Smart Manufacturing Innovation Institute): https://cesmii.org/
- DMDII (The Digital Manufacturing and Design Innovation Institute): http://dmdii.uilabs.org/
- IACMI (The Institute for Advanced Composites Manufacturing Innovation): http://iacmi.org/
- LIFT (Lightweight Innovations For Tomorrow): http://lift.technology/
- NextFlex: http://www.nextflex.us/
- NIIMBL (The National Institute for Innovation in Manufacturing Biopharmaceuticals): http://www.niimbl.us/
- Power America: http://www.poweramericainstitute.org/
- RAPID (Rapid Advancement in Process Intensification Deployment Institute): http://processintensification.org/
- REMADE (Reducing EMbodied-energy And Decreasing Emissions): http://www.rit.edu/gis/remade/index.html

***China***: Manufacturing industry in China increasingly faces persistent challenges from environmental issues, resource shortage, rising labor costs, and a slowdown in economic growth. As a result, the "Made in China 2025" strategy that began in 2015 to provide a 10-year action plan to radically transform the manufacturing sector. The goal is to turn the country from a quantity manufacturer to a high-end quality manufacturer. This strategy targets ten important areas that are vital for economic growth, i.e., information technology, aviation, railway equipment, power-grid, new materials, machinery, robotics, maritime equipment, energy-saving vehicles, and medical devices. Smart manufacturing is also identified as an opportunity for Chinese manufacturers to take the lead in the global competition. Three directions are highlighted to improve the "smartness" level of manufacturing, i.e., $i$) Developing new unmanned manufacturing systems with smart sensors, industrial robots, RFIDs, control systems, and automated production lines; $ii$) Realizing the internet-based information infrastructure to effectively and efficiently

* President's Council of Advisors on Science and Technology

coordinate the manufacturing network; $iii$) Developing industrial cloud platforms and big data analytical tools to help manufacturing enterprises make better decisions. The "Made in China 2025" strategy is seeking to promote data-driven innovation and smart technologies to pursue sustainable economic growth and upgrade China from the largest manufacturer in the world to a pioneering manufacturing power.

***European Union***: EU economy relies heavily on the manufacturing sector, which contributes 80% of all EU exports. However, the EU economic crisis has led to a decline of manufacturing throughput with more than 3.8 million jobs lost between 2009 and 2013. As such, EU Commission has organized several task forces to put together action plans to increase the competitiveness of EU manufacturing, including digitizing European industry, factory of the future, smart anything everywhere, and advanced manufacturing for clean production. Digital opportunities to make industry smarter that have been identified include the IoT, big data, artificial intelligence (AI), additive manufacturing (AM), robotics, and blockchain technologies. The "factory of the future" is a multi-year roadmap (2014-2020) to realize a smart factory that is clean, highly performing, environmental friendly and socially sustainable. The priority of the EU Commission is to digitalize the industry to make the best use of new technologies and manufacture high-quality digitalized products or service. A number of digital innovation hubs have also been established across Europe to help small, medium or large companies make the most of digital opportunities. The policies and strategies from the EU Commission are complemented and integrated by many national initiatives, for e.g.,

- **Germany**: Industry 4.0, Smart Service World, High-Tech Strategy 2020
- **Netherlands**: Smart Industry
- **Italy**: Internet of Things and Industry 4.0
- **Belgium**: Made Different, Marshall 4.0, Flanders make
- **France**: Alliance Industry of the Future, Industrie du Futur, Nouvelle France Industrielle
- **Spain**: Industria Conectada 4.0

In short, the EU Commission aims to lead a smooth transition to a smart economy, prepare to manufacture products & services of the next generation, improve innovation capacity across manufacturing industries, and increase the total Gross Domestic Product of the European Union.

In addition, United Kingdom announced the foresight project "Future of Manufacturing" in 2013 that provides the 2013-2050 strategical plan for the country to adapt to the megatrend of the global manufacturing revolution. This foresight project joins other initiatives such as High Value Manufacturing Catapult, Innovate UK, and EPSRC Manufacturing the Future to address key challenges on the UK manufacturing sector, e.g.,

- Adapt to increasing demands for personalized products and services
- The lack of highly skilled labor well trained in new technologies
- Sustainable manufacturing that is efficient in the use of materials and energy
- Digitalize manufacturing to realize the full potential of IoT sensing, big data analytics, intelligent systems, 3D printing, robotics, and new materials.

Furthermore, General Electric, Cisco, Intel, AT&T, and IBM founded the "Industrial Internet Consortium (IIC)" in 2014 to shape the future of industrial IoT systems. Currently, the IIC consists of more than 258 academic and industrial members who have invested heavily in IoT and CPS related projects. Thus far, the IIC has put together over 20 testbeds to demonstrate the implementation of IoT systems and data analytics to provide transformational outcomes in the industry. It is expected that IoT applications in manufacturing and factory settings will generate $1.2 to $3.7 trillion annually by 2025 [5]. Clearly, IoT and smart manufacturing will lead to significant economic and societal impacts.

# VII. IoT Challenges and Opportunities in Manufacturing

As the infrastructure of manufacturing systems become smarter, more and more operations are being carried out by an increasing number of machines. We observe that different machines may carry out the same or different functions or tasks, and some machines rely strongly on the output of other machines, just like a pipelined product line. Such strong connections may also vary dynamically depending on the different tasks being executed. In a word, the synergy of various machines has become critical for the overall performance of existing and future systems. The IoMT deploys a multitude of sensors to continuously monitor machine conditions. Sensor outputs, known as machine signatures, provide an unprecedented opportunity for optimal decision making in manufacturing. However, realizing the full potential of IoMT for smart manufacturing depends, to a great extent, on addressing the following challenges.

*The first challenge* is knowing the status of each machine. This status includes not only the fact of being busy or not, but also the health condition, in the sense of whether it is functioning properly or not. This status information is very important as it determines whether this machine can be counted on for task execution. The most straightforward method is to use sensors that can carry out both the sensing task and also provide some analysis based on signal processing of the sensed data. These sensors may be powered by wired supply or batteries. However, with the increasing number of machines, considering their expected lifetime of one to two decades, in some scenarios it is difficult to provide wired power or battery support. For example, wires limit the portability of sensors. Battery replacement is also sometimes challenging and time-consuming for these sensing systems. Also, batteries may not be safe or efficient in some extreme environments.

*The second challenge* is how to make use of the status of various machines to distribute tasks to each machine. Should each machine follow a strict static schedule? The potential malfunction of machines, and also dynamic changes in system-level tasks will result in schedule differently optimized towards energy, total operation time, etc. Under these circumstances, distributing tasks dynamically to machines based on the sensed status of each machine is key.

*The third challenge* originates from the communications between machines, possibly including the coordinating machines if they exist. It is noted that nowadays some tasks are carried out by machines from different sources, even from different countries. This reveals the possibility of problems in communication reliability and its impact on the collaboration between these machines. In addition, manufacturing assets are closed systems that cannot be controlled fully from the outside even if a two-way flow of information exists. Take a machine tool as an example. One can send G code to the machine to run it, but one cannot control the servos and spindles of the machines directly. This is yet another big challenge that must be overcome to enable full control and automation. Addressing these challenges will lead to new avenues of fundamental and applied research in manufacturing, sustainable manufacturing, IIoT and Cognitive Supply Chains.

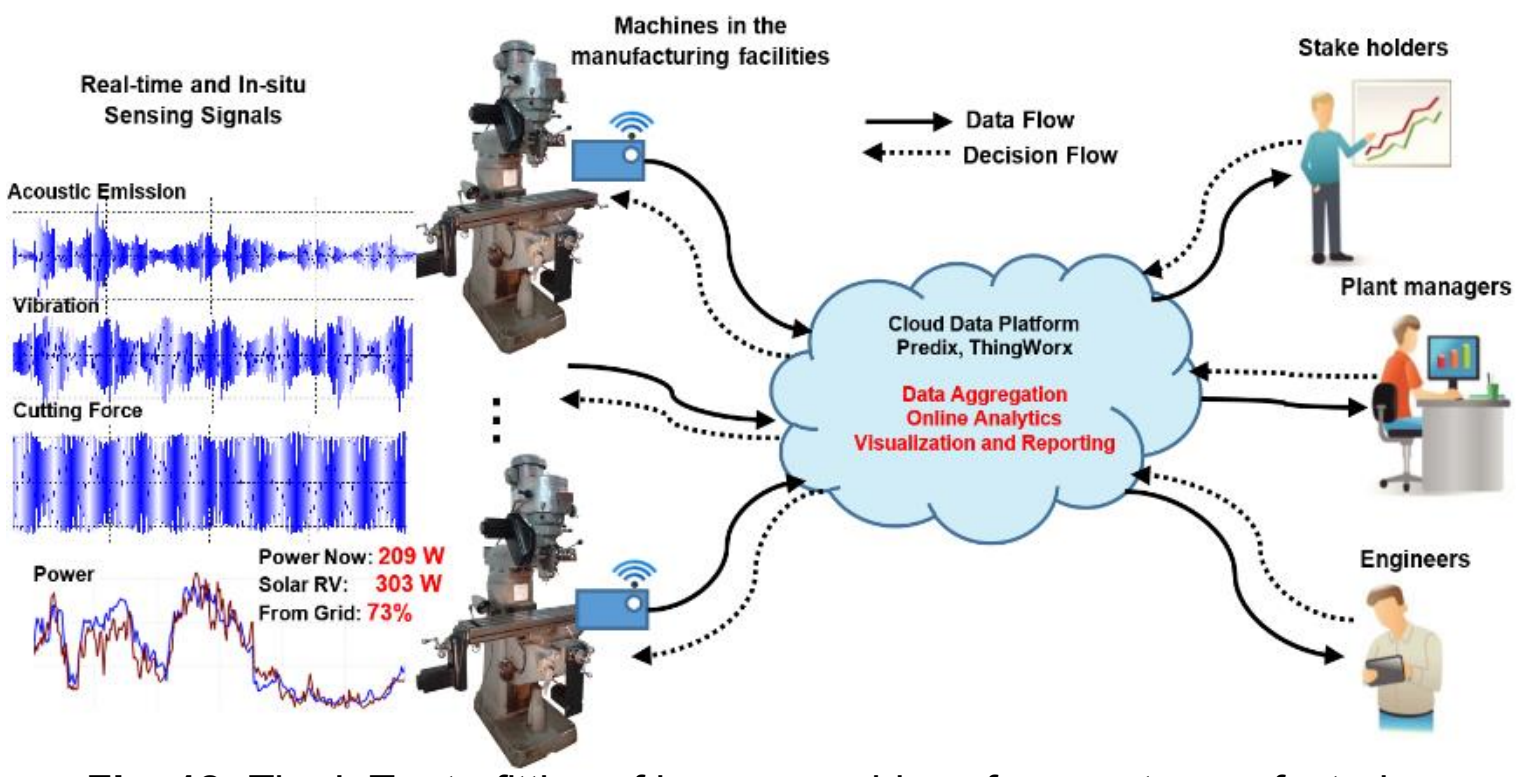

**Fig. 12:** The IoT retrofitting of legacy machines for smart manufacturing

**Opportunity 1. Retrofit legacy machines for smart manufacturing**

Although new manufacturing technologies and upstart companies arise, there are many existing manufacturing firms falling behind the wave of digital evolution. It is not uncommon that legacy (or old) machines exist in many small manufacturing firms. While legacy machines are invaluable assets for manufacturing firms and are fully utilized in production, they lack real-time and in-process sensing and control systems. As a result, small manufacturers increasingly lose competitive advantage in the global market because they are limited in information visibility and in the ability to cope with the greater complexity of modern manufacturing environments. Fundamental to the problem is establishing IoT connectivity between legacy machines. As shown in Fig. 12, IoMT sensing provides an unprecedented opportunity to retrofit legacy machines for digital manufacturing. As a result, there is an urgent need to develop new plug-and-play IoT sensors that continuously collect in-situ machine data, transmit the data to cloud storage, and communicate with other "things" and stake holders.

**Opportunity 2. Self-powered machine status sensing**

It is imperative to sense the status of a machine with a self-powered supply mechanism. By making the sensors self-powered, no wireline connection or battery is needed to provide a power supply. With additional wireless pairing and data transition, such a sensing system could be used efficiently in many machines, enhancing portability and reducing maintenance costs. Signal processing techniques, either preliminary raw data processing or end-to-end implementation of functions, could also be added to the sensor node.

**Opportunity 3. Machine service and tasks scheduling and distributing**

There is an opportunity to study optimized task distribution (scheduling) methodology between a group of machines for a set of tasks or services considering, in particular, the assistance of sensed machine data. For example, in a centralized system with a server center, there will the following questions to be answered: • When and what tasks should be distributed to which machines (in a dynamic distribution system)? • How to assess the potential contribution of a machine that is currently malfunctioning but may be fixed? How will this affect task scheduling and distribution policies? • What are the energy consumption and utilization of each machine?

**Opportunity 4. The synergy between IoMT machines**

It is also important to optimize the synergy between a set of machines collaborating remotely. Due to physical distance and unreliable message channels, there may be a temporary block in communication between machines in different places. What should a machine do when it finds itself isolated from other machines? How should the central coordinator be designed?

**Opportunity 5. Cloud computing and analytics**

The cloud data platform is a centralized data repository, which will include not only historical data collected from a large number of machines, but also on-line data from the machines *in-situ*. This data can be retrieved easily from the cloud platform to local computers to extract useful information and prototype algorithms that can be deployed in either the cloud or the IoT sensor devices. Data-driven system informatics and control is an indispensable step to the next generation of digital manufacturing. Cloud computing and analytics will open avenues of opportunity to optimize the management and planning of manufacturing operations, from quality management, power management, heat and cooling, sustainability and safety, to distribution and supply chain management.

**Opportunity 6. Blockchain enabled IoT**

IoMT things communicate with each other through the internet. Data security and privacy emerge as a big issue for the design, development, and deployment of IoT systems. Manufacturing industry is a part of the critical infrastructure of each country. Cyber-attacks on the IoMT system will directly disrupt manufacturing operations and other essential functions in pertinent industries. On the other hand, manufacturing is becoming global and distributed. IoMT things are not necessarily controlled by a centralized system. How to enable secure data sharing between IoMT things? Also, how to realize the decentralized control of IoMT things. One possible solution is the blockchain technology which is a distributed system managed by a peer-to-peer network to validate and ensure secure data transport using cryptography. Because blockchain offers an effective means of sharing data securely under decentralized control, it also provides a natural solution as a data sharing framework for IoMT systems. Though there is preliminary commercial work being done in this domain, more fundamental research is needed.

# VIII. Conclusions

To achieve competitive advantages in the global market, manufacturing industry is striving to create new products and services. As a result, advanced sensor technologies are used widely in manufacturing systems to increase information visibility and system controllability. Note that although sensors, data and IT systems may already be available in physical factories, they are not closely integrated up to the level of IoT. Recently, Industry 4.0 aims to boost the manufacturing system to a new generation of cyber-physical systems for smart manufacturing. IoT sensing collects enormous amounts of data from manufacturing systems in the physical world. Realizing the full potential of IoT for smart manufacturing requires new advances in analytical methodologies. The challenges now are “how to reflect physical manufacturing in cyberspace through data-driven information processing and modeling?” and “how to exploit the useful information and knowledge extracted from data to provide better manufacturing operations in the physical world?”

Indeed, smart manufacturing depends to a great extent on data-driven innovations to realize the seamless integration of cyber and physical spaces. Industry companies, trade groups, and standard organizations are racing against the clock to lead the evolution of Industry 4.0. A number of IoT architectures such as RAMI 4.0 and OPC UA have been proposed to define the communication structure of Industry 4.0. Note that RAMI 4.0 provides a reference architectural model to define 3 critical dimensions of manufacturing industry 4.0, i.e., Factory Hierarchy (i.e., product, field device, control device, station, work center, and enterprise), Architecture (i.e., Asset, Integration, communication, information, function, and business), and Product Life Cycle (i.e., from the initial design to the scrapyard). In addition, commercial IoT platforms such as GE Predix, ThingWorx, IBM Watson, Microsoft Azure, and Amazon AWS are readily available to enable physical "Things" and cyber-world applications to communicate and integrate with each other. The diverse types of IoT architectures and platforms are conducive to the acceleration of the development of IoT systems.

It may be noted that industry focuses more on the establishment of IoT standards and platforms, which help integrate existing sensors, IT and OT systems into the new IoT framework. There are many IoT case studies available from company websites for marketing purposes (also see Table III). However, IoT is still under development and faces technical issues for cyber-physical integration in the manufacturing system such as communication, big data, and control. For example, a single vibration sensor in the machine condition monitoring system generates data streams at high velocity. However, the cloud database has a limited bandwidth for data transmission and update frequency. Is it necessary to transmit all the raw data to the cloud, or just extract useful information for control decisions via embedded computing? In addition, data are "big" not just in terms of volume, but also in terms of variety and veracity. Note that there are a variety of data in manufacturing, from power profiles to machining parameters to acoustic emission to cutting force signals, each requiring a particular signal acquisition parameter. Also, veracity is particularly important in the IoT paradigm given the uncertainty (and the lack of quantification of uncertainty) of statistical models. Further, IoMT data analytics require manufacturing domain expertise to steer and gain value from the data. Most of commercial IoT platforms (e.g., Preidx, Azure) are not specifically designed and customized for the manufacturing industry, and are therefore limited in ability to fulfill the needs of smart manufacturing. In addition, the manufacturing industry is critical to national security. Cyber-attacks on manufacturing systems will impact the national economy and prosperity directly. Therefore, manufacturing assets are closed systems that cannot be fully controlled from the outside. A critical question is "how to enable secure data sharing and decentralized control of IoMT things?"

Manufacturing researchers have traditionally been less concerned about the issues of big data analytics, cybersecurity, cloud computing, system optimization in the large-scale IoT context. These research problems are critically important to improving the performance of manufacturing enterprises and achieving a high level of "smartness" in manufacturing. This paper presents a review of the development of IoT technologies and existing applications in manufacturing enterprises. Further, we provide a preliminary study to leverage IoMT and cloud computing to build virtual machine networks, thereby improving manufacturing decision-making capability through the cyber-physical integration of manufacturing enterprises. We hope our focused and limited review can serve as a catalyst to stimulate more in-depth and comprehensive studies that will focus on the development of novel IoMT technologies and analytical methodologies to improve manufacturing services and optimize manufacturing systems. Without a doubt, IoMT and smart manufacturing present a promising research paradigm with strong potential to revolutionize next-generation manufacturing enterprises.

## Acknowledgements

The authors would like to thank Chen Kan, Rui Zhu, Cheng-bang Chen, Bing Yao for their help in organizing and editing references used in this paper. Also, the authors thank Dr. Congbo Li for sharing the dataset of power profiles from machining opeartaions, as well as Dr. Yun Chen and Dr. Shijie Su for sharing the power profiles from welding operations. This work is supported by the National Science Foundation CAREER grant (CMMI-1617148). The author (HY) also thanks Lockheed Martin, Harold and Inge Marcus Career Professorship (HY) for additional financial support.

# Author Information

Hui Yang is the Harold and Inge Marcus Career Associate Professor in the Harold and Inge Marcus Department of Industrial and Manufacturing Engineering at The Pennsylvania State University, University Park, PA. His research interests are sensor-based modeling and analysis of complex systems for process monitoring, process control, system diagnostics, condition prognostics, quality improvement, and performance optimization. He received the NSF CAREER award in 2015, and multiple best paper awards from the international IEEE, IISE and INFORMS conferences. Dr. Yang is the president (2017-2018) of IISE Data Analytics and Information Systems Society, the president (2015-2016) of INFORMS Quality, Statistics and Reliability (QSR) society, and the program chair of 2016 Industrial and Systems Engineering Research Conference (ISERC). He is also an associate editor for IISE Transactions, IEEE Journal of Biomedical and Health Informatics (JBHI), and IEEE Robotics and Automation Letters (RA-L).

Soundar Kumara is the Allen, E., and Allen, M., Pearce Professor of Industrial and Manufacturing Engineering at Penn State. He holds an affiliate appointment with the school of Information Sciences and Technology. Dr. Kumara is a Fellow of Institute of Industrial Engineers (IIE), Fellow of the International Academy of Production Engineering (CIRP), Fellow of American Society of Mechanical Engineers (ASME), and a Fellow of the American Association for the Advancement of Science (AAAS). Kumara is a leader in industrial engineering for his pioneering and visionary interdisciplinary research in logistics and manufacturing. His unique approaches integrate mathematics, AI, pattern recognition, advanced computing, statistical physics and operations research, to solve problems in complex networks, product design and real time monitoring of manufacturing and logistics systems. He has laid the foundations of nonlinear dynamics based monitoring and diagnosis methodologies in manufacturing process monitoring. One of his papers on clustering in large networks in Physics Reviews – E is designated as a milestone paper for 2007, commemorating 25 years of PRE, which has published more than 50,000 articles since its beginning in 1993.

Satish T. S. Bukkapatnam serves as Rockwell International Professor with Department of Industrial and Systems Engineering department at Texas A&M University, College Station, TX, USA. He has previously served as an AT&T Professor at the Oklahoma State University and as an Assistant Professor at the University of Southern California. He is also the Director of Texas A&M Engineering Experimentation Station (TEES) Institute for Manufacturing Systems. He also holds an affiliate faculty appointment at Ecole Nationale Superior Arts et Metier (ENSAM), France. His research addresses the harnessing of high-resolution nonlinear dynamic information, especially from wireless MEMS sensors, to improve the monitoring and prognostics, mainly of ultraprecision and nanomanufacturing processes and machines, and cardiorespiratory processes. His research has led to 151 peer-reviewed publications (87 published/ accepted in journals and 64 in conference proceedings), five pending patents, 14 completed PhD dissertations, $5 million in grants as PI/Co-PI from the National Science Foundation, the U.S. Department of Defense, and the private sector, and 17 best-paper/poster recognitions. He is a fellow of the Institute for Industrial and Systems Engineers (IISE) and the Society of Manufacturing Engineers (SME), and he has been recognized with Oklahoma State University regents distinguished research, Halliburton outstanding college of engineering faculty, IISE Boeing technical innovation, IISE Eldin outstanding young industrial engineer, and SME Dougherty outstanding young manufacturing engineer awards. He currently serves as the editor of the IISE Transactions, Design and Manufacturing Focused Issue. He received his master's and Ph.D. degrees from the Pennsylvania State University and undergraduate degree from S.V. University, Tirupati, India.

Fugee Tsung is Professor of the Department of Industrial Engineering and Decision Analytics (IEDA), Director of the Quality and Data Analytics Lab, at the Hong Kong University of Science & Technology (HKUST), and Editor-in-Chief of the Journal of Quality Technology (JQT). He is a Fellow of the Institute of Industrial and Systems Engineers (IISE), Fellow of the American Society for Quality (ASQ), Fellow of the American Statistical Association (ASA), Academician of the International Academy for Quality (IAQ), and Fellow of the Hong Kong Institution of Engineers (HKIE). He received both his MSc and PhD from the University of Michigan, Ann Arbor and his BSc from National Taiwan University. He has authored over 100 refereed journal publications and is also the winner of the Best Paper Award for the IIE Transactions in 2003, 2009, 2017. His research interests include industrial big data and quality analytics.

# References

[1] U.S. Department of Commerce Bureau of Economic Analysis, http://www.bea.gov/industry/gdpbyind_data.htm.

[2] Smart Manufacturing Leadership Coalition, https://smartmanufacturingcoalition.org/.

[3] A. Kusiak, "Smart manufacturing," *Int J Prod Res,* vol. 56, no. 1-2, pp. 508-517, 2017.

[4] Telecommunication Development Bureau, International Telecommunication Union(ITU) ICT Facts and Figures 2017, http://www.itu.int/en/ITU-D/Statistics/Pages/facts/default.aspx.

[5] International Data Corp. http://www.idc.com/.

[6] K. Ashton. That 'Internet of Things' Thing, in the Real World, Things Matter More Than Ideas, http://www.rfidjournal.com/articles/view?4986.

[7] J. M. Govardhan, S. T. S. Bukkapatnam, Y. Bhamare, P. K. Rao and V. Rajamani, "Statistical analysis and design of RFID systems for monitoring vehicle ingress/egress in warehouse environments," *International Journal of Radio Frequency Identification Technology and Applications,* vol. 1, no. 2, pp. 123-146, 2007.

[8] J. Zhou and J. Shi, "RFID localization algorithms and applications-a review," *Journal of Intelligent Manufacturing,* vol. 20, no. 6, pp. 695, 2008.

[9] L. D. Xu, W. He and S. Li, "Internet of Things in Industries: A Survey," *IEEE Transactions on Industrial Informatics,* vol. 10, no. 4, pp. 2233-2243, 2014.

[10] C. Ok, S. Lee, P. Mitra and S. Kumara. Distributed energy balanced routing for wireless sensor networks. *Computers & Industrial Engineering 57(1),* pp. 125-135, 2009.

[11] I. F. Akyildiz, W. Su, Y. Sankarasubramaniam and E. Cayirci, "Wireless sensor networks: a survey," *Computer Networks,* vol. 38, no. 4, pp. 393-422, 2002.

[12] C. A. Tokognon, B. Gao, G. Y. Tian and Y. Yan, "Structural Health Monitoring Framework Based on Internet of Things: A Survey," *IEEE Internet of Things Journal,* vol. 4, no. 3, pp. 619-635, 2017.

[13] X. Yang and L. Chen, "Using multi-temporal remote sensor imagery to detect earthquake-triggered landslides," *International Journal of Applied Earth Observation and Geoinformation,* vol. 12, no. 6, pp. 487-495, 2010.

[14] H. Ren, J. Long, Z. Gao and P. Orenstein, "Passenger Assignment Model Based on Common Route in Congested Transit Networks," *Journal of Transportation Engineering,* vol. 138, no. 12, pp. 1484-1494, 2012.

[15] P. Rao, S. Bukkapatnam, O. Beyca, Z. Kong and R. Komanduri, "Real-Time Identification of Incipient Surface Morphology Variations in Ultraprecision Machining Process," *Journal of Manufacturing Science and Engineering,* vol. 136, no. 2, pp. 021008-021008-11, 2014.

[16] O. F. Beyca, P. K. Rao, Z. Kong, S. T. S. Bukkapatnam and R. Komanduri, "Heterogeneous Sensor Data Fusion Approach for Real-time Monitoring in Ultraprecision Machining (UPM) Process Using Non-Parametric Bayesian Clustering and Evidence Theory," *IEEE Transactions on Automation Science and Engineering,* vol. 13, no. 2, pp. 1033-1044, 2016.

[17] A. Kamilaris and A. Pitsillides, "Mobile Phone Computing and the Internet of Things: A Survey," *IEEE Internet of Things Journal,* vol. 3, no. 6, pp. 885-898, 2016.

[18] A. Pantelopoulos and N. G. Bourbakis, "A Survey on Wearable Sensor-Based Systems for Health Monitoring and Prognosis," *IEEE Transactions on Systems, Man, and Cybernetics, Part C (Applications and Reviews),* vol. 40, no. 1, pp. 1-12, 2010.

[19] A. Sa-ngasoongsong, J. Kunthong, V. Sarangan, X. Cai and S. T. S. Bukkapatnam, "A Low-Cost, Portable, High-Throughput Wireless Sensor System for Phonocardiography Applications," *Sensors,* vol. 12, no. 8, pp. 10851-10870, 2012.

[20] C. Gomez and J. Paradells, "Wireless home automation networks: A survey of architectures and technologies," *IEEE Communications Magazine,* vol. 48, no. 6, pp. 92-101, 2010.

[21] G. W. Tan, K. Ooi, S. Chong and T. Hew, "NFC mobile credit card: The next frontier of mobile payment?" *Telematics and Informatics,* vol. 31, no. 2, pp. 292-307, 2014.

[22] R. Sanchez-Iborra and M. Cano, "State of the Art in LP-WAN Solutions for Industrial IoT Services," *Sensors,* vol. 16, no. 5, pp. 708-721, 2016.

[23] Neul, http://www.neul.com/neul/.

[24] R. S. Sinha, Y. Wei and S. Hwang, "A survey on LPWA technology: LoRa and NB-IoT," *ICT Express,* vol. 3, no. 1, pp. 14-21, 2017.

[25] S. L. Levin and S. Schmidt, "IPv4 to IPv6: Challenges, solutions, and lessons," *Telecommunications Policy,* vol. 38, no. 11, pp. 1059-1068, 2014.

[26] X. Wang, H. Chen and D. Le, "A novel IPv6 address configuration for a 6LoWPAN-based WBAN," *Journal of Network and Computer Applications,* vol. 61, pp. 33-45, 2016.

[27] B. Edrington, B. Zhao, A. Hansel, M. Mori and M. Fujishima, "Machine Monitoring System Based on MTConnect Technology," *Procedia CIRP,* vol. 22, pp. 92-97, 2014.

[28] P. Lei, L. Zheng, C. Li and X. Li, "MTConnect Enabled Interoperable Monitoring System for Finish Machining Assembly Interfaces of Large-scale Components," *Procedia CIRP,* vol. 56, pp. 378-383, 2016.

[29] H. Zimmermann, "OSI Reference Model-The ISO Model of Architecture for Open Systems Interconnection," *Communications, IEEE Transactions on,* vol. 28, no. 4, pp. 425-432, 1980.

[30] R. Y. Zhong, L. Wang and X. Xu, "An IoT-enabled Real-time Machine Status Monitoring Approach for Cloud Manufacturing," *Procedia CIRP,* vol. 63, pp. 709-714, 2017.

[31] S. S. Manvi and G. Krishna Shyam, "Resource management for Infrastructure as a Service (IaaS) in cloud computing: A survey," *Journal of Network and Computer Applications,* vol. 41, pp. 424-440, 2014.

[32] A. J. Ferrer, D. G. Pérez and R. S. González, "Multi-cloud Platform-as-a-service Model, Functionalities and Approaches," *Procedia Computer Science,* vol. 97, pp. 63-72, 2016.

[33] A. Amiri, "Application placement and backup service in computer clustering in Software as a Service (SaaS) networks," *Computers & Operations Research,* vol. 69, pp. 48-55, 2016.

[34] S. Jayaram, H. I. Connacher and K. W. Lyons, "Virtual assembly using virtual reality techniques," *Computer-Aided Design,* vol. 29, no. 8, pp. 575-584, 1997.

[35] A. Y. C. Nee, S. K. Ong, G. Chryssolouris and D. Mourtzis, "Augmented reality applications in design and manufacturing," *CIRP Annals - Manufacturing Technology,* vol. 61, no. 2, pp. 657-679, 2012.

[36] J. Lee, H. Kao and S. Yang, "Service Innovation and Smart Analytics for Industry 4.0 and Big Data Environment," *Procedia CIRP,* vol. 16, pp. 3-8, 2014.

[37] R. Y. Zhong, C. Xu, C. Chen and G. Q. Huang, "Big Data Analytics for Physical Internet-based intelligent manufacturing shop floors," *International Journal of Production Research,* vol. 55, no. 9, pp. 2610-2621, 2017.

[38] S. T. S. Bukkapatnam, A. Lakhtakia and S. R. T. Kumara, "Analysis of sensor signals shows turning on a lathe exhibits low-dimensional chaos," *Physical Review E,* vol. 52, no. 3, pp. 2375-2387, 1995.

[39] C. Cheng, A. Sa-Ngasoongsong, O. Beyca, T. Le, H. Yang, Z. Kong and S. T. S. Bukkapatnam, "Time series forecasting for nonlinear and non-stationary processes: a review and comparative study," *IIE Transactions,* vol. 47, no. 10, pp. 1053-1071, 2015.

[40] A. Kusiak, "Smart manufacturing must embrace big data," *Nature,* vol. 544, no. 7648, pp. 23–25, 2017.

[41] F. Tao, Q. Qi, A. Liu and A. Kusiak. Data-driven smart manufacturing. *Journal of Manufacturing Systems* 2018, .

[42] A. Kumar, R. Shankar, A. Choudhary and L. S. Thakur, "A big data MapReduce framework for fault diagnosis in cloud-based manufacturing," *International Journal of Production Research,* vol. 54, no. 23, pp. 7060-7073, 2016.

[43] Y. Bao, L. Ren, L. Zhang, X. Zhang and Y. Luo, "Massive sensor data management framework in cloud manufacturing based on hadoop," in *IEEE 10th International Conference on Industrial Informatics,* 2012, pp. 397-401.

[44] d. U. Saenz, A. Artiba and R. Pellerin, "Manufacturing execution system – a literature review," *Production Planning & Control,* vol. 20, no. 6, pp. 525-539, 2009.

[45] M. Quiescenti, M. Bruccoleri, U. La Commare, N. L. Diega and G. Perrone, "Business process-oriented design of Enterprise Resource Planning (ERP) systems for small and medium enterprises," *International Journal of Production Research,* vol. 44, no. 18-19, pp. 3797-3811, 2006.

[46] S. R. T. Kumara and S. T. S. Bukkapatnam, "Characterization and monitoring of nonlinear dynamics and chaos in manufacturing enterprise systems," in *Network Science, Nonlinear Science and Infrastructure Systems*, T. Friesz, Ed. 2007, pp.99-122.

[47] H. Yang, S. T. S. Bukkapatnam and R. Komanduri, "Nonlinear adaptive wavelet analysis of electrocardiogram signals," *Physical Review E,* vol. 76, no. 2, pp. 026214, 2007.

[48] S. T. Bukkapatnam, S. R. Kumara and A. Lakhtakia, "Fractal estimation of flank wear in turning," *Journal of Dynamic Systems, Measurement, and Control,* vol. 122, no. 1, pp. 89-94, 2000.

[49] S. T. S. Bukkapatnam, S. R. Kumara and A. Lakhtakia, "Local eigenfunctions-based suboptimal wavelet packet representation of contaminated chaotic signals," *IMA Journal of Applied Mathematics,* vol. 63, pp. 149-160, 1999.

[50] S. T. S. Bukkapatnam, S. Kumara and A. Lakhtakia, "Analysis of acoustic emission signals in machining," *ASME Transactions on Manufacturing Science and Engineering,* vol. 121, pp. 568-573, 1999.

[51] C. Koh, J. Shi and W. Williams, "Tonnage Signature Analysis Using the Orthogonal (HAAR) Transforms," *Transactions of NAMRI/SME,* vol. 23, no. 1, pp. 229-234, 1995.

[52] J. Jin and J. Shi, "Feature-Preserving Data Compression of Stamping Tonnage Information Using Wavelets," *Technometrics,* vol. 41, no. 4, pp. 327-339, 1999.

[53] J. Jin and J. Shi, "Diagnostic Feature Extraction from Stamping Tonnage Signals Based on Design of Experiment," *ASME Transactions, Journal of Manufacturing Science and Engineering,* vol. 122, no. 2, pp. 360-369, 2000.

[54] Y. Ding, L. Zeng and S. Zhou, "Phase I analysis for monitoring nonlinear profiles in manufacturing processes," *Journal of Quality Technology,* vol. 38, no. 3, pp. 199-216, 2006.

[55] Bukkapatnam, S., S. Kumara, A. Lakhtakia,and P.Srinivasan, "Neighborhood method and its coupling with the wavelet method for nonlinear signal separation of contaminated chaotic time-series data," *Signal Processing,* vol. 82, pp. 1351-1374, 2002.

[56] S. Bukkapatnam, H. Yang and F. Modhavi, "Towards prediction of nonlinear and nonstationary evolution of customer preferences using local markov models," in *The Art and Science Behind Successful Product Launches*, N. R. Srinivasa Raghavan and John A. Cafeo, Eds. Springer, 2009, pp.271-287.

[57] H. Yang and Y. Chen, "Heterogeneous recurrence monitoring and control of nonlinear stochastic processes," *Chaos,* vol. 24, no. 1, pp. 013138, 2014.

[58] Y. Chen and H. Yang, "Heterogeneous recurrence representation and quantification of dynamic transitions in continuous nonlinear processes," *The European Physical Journal B,* vol. 89, no. 6, pp. 155, 2016.

[59] C. Cheng, C. Kan and H. Yang, "Heterogeneous recurrence analysis of heartbeat dynamics for the identification of sleep apnea events," *Computers in Biology and Medicine,* vol. 75, pp. 10-18, 2016.

[60] C. Kan, C. Cheng and H. Yang, "Heterogeneous recurrence monitoring of dynamic transients in ultraprecision machining processes," *Journal of Manufacturing Systems,* vol. 41, pp. 178-187, 2016.

[61] G. Liu and H. Yang, "Self-organizing network for group variable selection and predictive modeling," *Annals of Operation Research,* pp. 1-22, 2017.

[62] H. Yang, "Multiscale Recurrence Quantification Analysis of Spatial Cardiac Vectorcardiogram (VCG) Signals," *Biomedical Engineering, IEEE Transactions on,* vol. 58, no. 2, pp. 339-347, 2011.

[63] Y. Chen and H. Yang, "Multiscale recurrence analysis of long-term nonlinear and nonstationary time series," *Chaos, Solitons & Fractals,* vol. 45, no. 7, pp. 978-987, 2012.

[64] H. Yang, S. T. S. Bukkapatnam and L. G. Barajas, "Local recurrence based performance prediction and prognostics in the nonlinear and nonstationary systems," *Pattern Recognit,* vol. 44, no. 8, pp. 1834-1840, 2011.

[65] S. T. S. Bukkapatnam and C. Cheng, "Forecasting the evolution of nonlinear and nonstationary systems using recurrence-based local Gaussian process models," *Physical Review E,* vol. 82, no. 5, pp. 056206, 2010.

[66] J. Li and J. Shi, "Knowledge discovery from observational data for process control using causal bayesian networks," *IIE Transactions,* vol. 39, no. 6, pp. 681-690, 2007.

[67] X. Zhang and Q. Huang, "Analysis of interaction structure among multiple functional process variables for process control in semiconductor manufacturing," *Semiconductor Manufacturing, IEEE Transactions on,* vol. 23, no. 2, pp. 263-272, 2010.

[68] J. Shi, *Stream of Variation Modeling and Analysis for Multistage Manufacturing Processes.* CRC Press, Taylor & Francis Group, 2006.

[69] J. Liu, J. Shi and S. J. Hu, "Quality assured setup planning based on the stream of variation model for multi-stage machining processes," *IIE Transactions, Quality and Reliability Engineering,* vol. 41, pp. 323-334, 2009.

[70] A. J. J. Braaksma, A. J. Meesters, W. Klingenberg and C. Hicks, "A quantitative method for Failure Mode and Effects Analysis," *International Journal of Production Research,* vol. 50, no. 23, pp. 6904-6917, 2012.

[71] N. Gebraeel, "Sensory-Updated Residual Life Distributions for Components With Exponential Degradation Patterns," *Automation Science and Engineering, IEEE Transactions on,* vol. 3, no. 4, pp. 382-393, 2006.

[72] L. Bian, N. Gebraeel and J. P. Kharoufeh, "Degradation modeling for real-time estimation of residual lifetimes in dynamic environments," *IIE Transactions,* vol. 47, no. 5, pp. 471-486, 2015.

[73] H. Yang, S. T. S. Bukkapatnam and L. G. Barajas, "Continuous flow modelling of multistage assembly line system dynamics," *International Journal of Computer Integrated Manufacturing,* vol. 26, no. 5, pp. 401-411, 2013.

[74] M. C. Fu, "Optimization via simulation: A review," *Annals of Operations Research,* vol. 53, no. 1, pp. 199-247, 1994.

[75] F. Tao, Y. Cheng, L. D. Xu, L. Zhang and B. H. Li, "CCIoT-CMfg: Cloud Computing and Internet of Things-Based Cloud Manufacturing Service System," *IEEE Transactions on Industrial Informatics,* vol. 10, no. 2, pp. 1435-1442, 2014.

[76] D. Georgakopoulos, P. P. Jayaraman, M. Fazia, M. Villari and R. Ranjan, "Internet of Things and Edge Cloud Computing Roadmap for Manufacturing," *IEEE Cloud Computing,* vol. 3, no. 4, pp. 66-73, 2016.

[77] Y. C. Lin, M. H. Hung, H. C. Huang, C. C. Chen, H. C. Yang, Y. S. Hsieh and F. T. Cheng, "Development of Advanced Manufacturing Cloud of Things (AMCoT)—A Smart Manufacturing Platform," *IEEE Robotics and Automation Letters,* vol. 2, no. 3, pp. 1809-1816, 2017.

[78] Y. Zhang, G. Zhang, J. Wang, S. Sun, S. Si and T. Yang, "Real-time information capturing and integration framework of the internet of manufacturing things," *Int. J. Comput. Integr. Manuf.,* vol. 28, no. 8, pp. 811-822, 2015.

[79] L. Monostori, B. Kádár, T. Bauernhansl, S. Kondoh, S. Kumara, G. Reinhart, O. Sauer, G. Schuh, W. Sihn and K. Ueda, "Cyber-physical systems in manufacturing," *CIRP Annals - Manufacturing Technology,* in print 2016.

[80] K. Thramboulidis and F. Christoulakis, "UML4IoT—A UML-based approach to exploit IoT in cyber-physical manufacturing systems," *Computers in Industry,* vol. 82, pp. 259-272, 2016.

[81] F. Tao, J. Cheng and Q. Qi, "IIHub: an Industrial Internet-of-Things Hub Towards Smart Manufacturing Based on Cyber-Physical System," *IEEE Transactions on Industrial Informatics, PP(99),* pp. 1-1, 2017, .

[82] F. Tao, D. Zhao, Y. Hu and Z. Zhou, "Resource Service Composition and Its Optimal-Selection Based on Particle Swarm Optimization in Manufacturing Grid System," *IEEE Transactions on Industrial Informatics,* vol. 4, no. 4, pp. 315-327, 2008.

[83] F. Tao and Q. Qi, "New IT Driven Service-Oriented Smart Manufacturing: Framework and Characteristics," *IEEE Transactions on Systems, Man, and Cybernetics: Systems, PP(99),* pp. 1-11, 2017, .

[84] R. F. Babiceanu and R. Seker, "Big Data and virtualization for manufacturing cyber-physical systems: A survey of the current status and future outlook," *Computers in Industry,* vol. 81, pp. 128-137, 2016.

[85] G. Adamson, L. Wang and P. Moore, "Feature-based control and information framework for adaptive and distributed manufacturing in cyber physical systems," *Journal of Manufacturing Systems,* vol. 43, pp. 305-315, 2017.

[86] J. Qin, Y. Liu and R. Grosvenor, "A Framework of Energy Consumption Modelling for Additive Manufacturing Using Internet of Things," *Procedia CIRP,* vol. 63, pp. 307-312, 2017.

[87] Y. S. Tan, Y. T. Ng and J. S. C. Low, "Internet-of-Things Enabled Real-time Monitoring of Energy Efficiency on Manufacturing Shop Floors," *Procedia CIRP,* vol. 61, pp. 376-381, 2017.

[88] F. K. Shaikh, S. Zeadally and E. Exposito, "Enabling Technologies for Green Internet of Things," *IEEE Systems Journal,* vol. 11, no. 2, pp. 983-994, 2017.

[89] F. Tao, Y. Zuo, L. D. Xu, L. Lv and L. Zhang, "Internet of Things and BOM-Based Life Cycle Assessment of Energy-Saving and Emission-Reduction of Products," *IEEE Transactions on Industrial Informatics,* vol. 10, no. 2, pp. 1252-1261, 2014.

[90] A. Rymaszewska, P. Helo and A. Gunasekaran, "IoT powered servitization of manufacturing – an exploratory case study," *International Journal of Production Economics,* in press, 2017.

[91] D. Li, M. Ren and G. Meng, "Application of Internet of Things Technology on Predictive Maintenance System of Coal Equipment," *Procedia Engineering,* vol. 174, pp. 885-889, 2017.

[92] Y. Xu and M. Chen, "Improving Just-in-Time Manufacturing Operations by Using Internet of Things Based Solutions," *Procedia CIRP,* vol. 56, pp. 326-331, 2016.

[93] Y. Ding, P. Kim, D. Ceglarek and J. Jin, "Optimal sensor distribution for variation diagnosis in multistation assembly processes," *IEEE Transactions on Robotics and Automation,* vol. 19, no. 4, pp. 543-556, 2003.

[94] Y. Ding, E. A. Elsayed, S. Kumara, J. C. Lu, F. Niu and J. Shi, "Distributed Sensing for Quality and Productivity Improvements," *IEEE Transactions on Automation Science and Engineering,* vol. 3, no. 4, pp. 344-359, 2006.

[95] D. Boos, H. Guenter, G. Grote and K. Kinder, "Controllable accountabilities: the Internet of Things and its challenges for organisations," *Behaviour & Information Technology,* vol. 32, no. 5, pp. 449-467, 2013.

[96] E. Sun, X. Zhang and Z. Li, "The internet of things (IOT) and cloud computing (CC) based tailings dam monitoring and pre-alarm system in mines," *Safety Science,* vol. 50, no. 4, pp. 811-815, 2012.

[97] D. Podgórski, K. Majchrzycka, A. Dąbrowska, G. Gralewicz and M. Okrasa, "Towards a conceptual framework of OSH risk management in smart working environments based on smart PPE, ambient intelligence and the Internet of Things technologies," *International Journal of Occupational Safety and Ergonomics,* vol. 23, no. 1, pp. 1-20, 2017.

[98] B. Guo, D. Zhang, Z. Wang, Z. Yu and X. Zhou, "Opportunistic IoT: Exploring the harmonious interaction between human and the internet of things," *Journal of Network and Computer Applications,* vol. 36, no. 6, pp. 1531-1539, 2013.

[99] A. A. Nazari Shirehjini and A. Semsar, "Human interaction with IoT-based smart environments," *Multimedia Tools and Applications,* vol. 76, no. 11, pp. 13343-13365, 2017.

[100] T. Cheng, G. C. Migliaccio, T. Jochen and U. C. Gatti, "Data Fusion of Real-Time Location Sensing and Physiological Status Monitoring for Ergonomics Analysis of Construction Workers," *Journal of Computing in Civil Engineering,* vol. 27, no. 3, pp. 320-335, 2013.

[101] R. D. Meller, *Functional Design of Physical Internet Facilities: A Road-Based Transit Center.* Faculté des sciences de l'administration, Université Laval, 2013.

[102] Y. Cheng, F. Tao, L. Xu and D. Zhao, "Advanced manufacturing systems: supply–demand matching of manufacturing resource based on complex networks and Internet of Things," *Enterprise Information Systems,* pp. 1-18, 2016.

[103] P. J. Reaidy, A. Gunasekaran and A. Spalanzani, "Bottom-up approach based on Internet of Things for order fulfillment in a collaborative warehousing environment," *International Journal of Production Economics,* vol. 159, pp. 29-40, 2015.

[104] T. Qu, M. Thürer, J. Wang, Z. Wang, H. Fu, C. Li and G. Q. Huang, "System dynamics analysis for an Internet-of-Things-enabled production logistics system," *International Journal of Production Research,* vol. 55, no. 9, pp. 2622-2649, 2017.

[105] T. Fan, F. Tao, S. Deng and S. Li, "Impact of RFID technology on supply chain decisions with inventory inaccuracies," *International Journal of Production Economics,* vol. 159, pp. 117-125, 2015.

[106] G. Hwang, J. Lee, J. Park and T. Chang, "Developing performance measurement system for Internet of Things and smart factory environment," *International Journal of Production Research,* vol. 55, no. 9, pp. 2590-2602, 2017.

[107] L. Zhou, A. Y. L. Chong and E. W. T. Ngai, "Supply chain management in the era of the internet of things," *International Journal of Production Economics,* vol. 159, pp. 1-3, 2015.

[108] G. E. P. Box and R. D. Meyer, "An Analysis for Unreplicated Fractional Factorials," *Technometrics,* vol. 28, no. 1, pp. 11-18, 1986.

[109] Y. Chen and H. Yang, "A novel information-theoretic approach for variable clustering and predictive modeling using Dirichlet process mixtures," *Scientific Reports,* vol. 6, no. 38913, pp. 1-13, 2016.

[110] S. Zhou, B. Sun and J. Shi, "An SPC monitoring system for cycle-based waveform signals using haar transform," *Automation Science and Engineering, IEEE Transactions on,* vol. 3, no. 1, pp. 60-72, 2006.

[111] H. Yang, C. Kan, Y. Chen and G. Liu, "Spatiotemporal differentiation of myocardial infarctions," *IEEE Transactions on Automation Science and Engineering,* vol. 10, no. 4, pp. 938-947, 2013.

[112] A. M. Bronstein, M. M. Bronstein and R. Kimmel, "Generalized multidimensional scaling: A framework for isometry-invariant partial surface matching," *Proceedings of the National Academy of Sciences of the United States of America,* vol. 103, no. 5, pp. 1168-1172, 2006.

[113] Groenen, P., van de Velden,M., "Multidimensional Scaling by Majorization: A Review," *Journal of Statistical Software,* vol. 73, no. 8, pp. 1-26, 2016.

[114] C. Kan, H. Yang and S. Kumara, "Parallel Computing and Network Analytics for Fast Internet-of-Things (IOT) Machine Information Processing and Condition Monitoring," *Journal of Manufacturing Systems,* vol. 46, pp. 282-293, 2018.

[115] G. E. Dower and H. B. Machado, "XYZ data interpreted by a 12-lead computer program using the derived electrocardiogram," *Journal of Electrocardiology,* vol. 12, no. 3, pp. 249-261, 1979.

[116] K. P. Oliveira-Esquerre, D. E. Seborg, M. Mori and R. E. Bruns, "Application of steady-state and dynamic modeling for the prediction of the BOD of an aerated lagoon at a pulp and paper mill: Part II. Nonlinear approaches," *Chemical Engineering Journal,* vol. 105, no. 1-2, pp. 61-69, 2004.

[117] Y. Chen and H. Yang, "Self-organized neural network for the quality control of 12-lead ECG signals," *Physiological Measurement,* vol. 33, no. 9, pp. 1399, 2012.

[118] H. Yang and F. Leonelli, "Self-organizing visualization and pattern matching of vectorcardiographic QRS waveforms," *Computers in Biology and Medicine,* vol. 79, pp. 1-9, 2016.

[119] Y. Chen and H. Yang, "Sparse Modeling and Recursive Prediction of Space–Time Dynamics in Stochastic Sensor Networks," *Automation Science and Engineering, IEEE Transactions on,* vol. 13, no. 1, pp. 215-226, 2016.

[120] A. Clauset, C. Shalizi and M. Newman, "Power-Law Distributions in Empirical Data," *SIAM Review,* vol. 51, no. 4, pp. 661-703, 2009.

[121] R. Albert and A. -. Barabási, "Statistical mechanics of complex networks," *Reviews of Modern Physics,* vol. 74, no. 1, pp. 47-97, 2002.

[122] U. N. Raghavan, R. Albert and S. Kumara, "Near linear time algorithm to detect community structures in large-scale networks," *Physical Review E,* vol. 76, no. 3, pp. 036106, 2007.

[123] L. Cui, S. Kumara and R. Albert, "Complex Networks: An Engineering View," *IEEE Circuits and Systems Magazine,* vol. 10, no. 3, pp. 10-25, 2010.

[124] H. Yang and G. Liu, "Self-organized topology of recurrence-based complex networks," *Chaos,* vol. 23, no. 4, pp. 043116, 2013.

[125] NIST Cybersecurity Framework, https://www.nist.gov/cyberframework.

[126] M. J. Hutchins, R. Bhinge, M. K. Micali, S. L. Robinson, J. W. Sutherland and D. Dornfeld, "Framework for Identifying Cybersecurity Risks in Manufacturing," *Procedia Manufacturing,* vol. 1, pp. 47-63, 2015.

[127] Z. DeSmit, A. E. Elhabashy, L. J. Wells and J. A. Camelio, "An approach to cyber-physical vulnerability assessment for intelligent manufacturing systems," *Journal of Manufacturing Systems,* vol. 43, pp. 339-351, 2017.

[128] L. D. Sturm, C. B. Williams, J. A. Camelio, J. White and R. Parker, "Cyber-physical vulnerabilities in additive manufacturing systems: A case study attack on the .STL file with human subjects," *Journal of Manufacturing Systems,* vol. 44, pp. 154-164, 2017.

[129] L. Duan, D. Liu, Y. Zhang, S. Chen, R. P. Liu, B. Cheng and J. Chen, "Secure Data-Centric Access Control for Smart Grid Services Based on Publish/Subscribe Systems," *ACM Transactions on Internet Technology,* vol. 16, no. 4, pp. 23:1-23:17, 2016.

[130] S. Seo, J. Won and E. Bertino, "pCLSC-TKEM: A Pairing-free Certificateless Signcryption-tag Key Encapsulation Mechanism for a Privacy-Preserving IoT," *Transactions on Data Privacy,* vol. 9, no. 2, pp. 101-130, 2016.

[131] Y. Zhang, Y. Xiang and X. Huang, "Password-Authenticated Group Key Exchange: A Cross-Layer Design," *ACM Transactions on Internet Technology,* vol. 16, no. 4, pp. 24:1-24:20, 2016.

[132] N. Saxena, S. Grijalva and N. S. Chaudhari, "Authentication Protocol for an IoT-Enabled LTE Network," *ACM Transactions on Internet Technology,* vol. 16, no. 4, pp. 25:1-25:20, 2016.

[133] S. Siboni, A. Shabtai, N. O. Tippenhauer, J. Lee and Y. Elovici, "Advanced Security Testbed Framework for Wearable IoT Devices," *ACM Transactions on Internet Technology,* vol. 16, no. 4, pp. 26:1-26:25, 2016.

[134] A. Saeed, A. Ahmadinia, A. Javed and H. Larijani, "Random Neural Network Based Intelligent Intrusion Detection for Wireless Sensor Networks," *Procedia Computer Science,* vol. 80, pp. 2372-2376, 2016.

[135] H. Vincent, L. Wells, P. Tarazaga and J. Camelio, "Trojan Detection and Side-channel Analyses for Cyber-security in Cyber-physical Manufacturing Systems," *Procedia Manufacturing,* vol. 1, pp. 77-85, 2015.

[136] L. Thames and D. Schaefer, "Cybersecurity for industry 4.0 and advanced manufacturing environments with ensemble intelligence," in *Cybersecurity for Industry 4.0: Analysis for Design and Manufacturing*, L. Thames and D. Schaefer, Eds. Cham: Springer International Publishing, 2017, pp.243-265.

[137] N. Z. Aitzhan and D. Svetinovic, "Security and Privacy in Decentralized Energy Trading through Multi-signatures, Blockchain and Anonymous Messaging Streams," *IEEE Transactions on Dependable and Secure Computing,* vol. PP, no. 99, pp. 1-1, 2016.

[138] K. Christidis and M. Devetsikiotis, "Blockchains and Smart Contracts for the Internet of Things," *IEEE Access,* vol. 4, pp. 2292-2303, 2016.

[139] P. Ghuli, U. P. Kumar and R. Shettar, "A review on blockchain application for decentralized decision of ownership of IoT devices," *Advances in Computational Sciences and Technology,* vol. 10, no. 8, pp. 2449-2456, 2017.

[140] A. Bahga and V. K. Madisetti, "Blockchain Platform for Industrial Internet of Things," *Journal of Software Engineering and Applications,* vol. 9, no. 10, pp. 533-546, 2016.

[141] President's Council of Advisors on Science and Technology, "Report to the president on ensuring american leadership in advanced manufacturing," 2012.